\documentclass[twocolumn,amsmath]{aastex62}

\newcommand{\vect}[1]{\mathbf{#1}}   

\usepackage{soul}

\received{January 1, 2018}
\revised{January 7, 2018}
\accepted{\today}
\submitjournal{ApJ}

\shorttitle{Energetic particles at TRAPPIST-1}
\shortauthors{Fraschetti, Drake et al.}

\begin{document}

\title{Stellar energetic particles in the magnetically turbulent habitable zones of TRAPPIST-1-like planetary systems}

\correspondingauthor{Federico Fraschetti}
\email{federico.fraschetti@cfa.harvard.edu}

\author{F. Fraschetti}
\affiliation{Center for Astrophysics $|$ Harvard \& Smithsonian, Cambridge, MA, 02138, USA; Dept. of Planetary Sciences-Lunar and Planetary Laboratory, University of Arizona, Tucson, AZ, 85721, USA}

\author{J. J. Drake}
\affiliation{Center for Astrophysics $|$ Harvard \& Smithsonian, Cambridge, MA, 02138, USA} 

\author{J. D. Alvarado-G\'omez}
\affiliation{Center for Astrophysics $|$ Harvard \& Smithsonian, Cambridge, MA, 02138, USA} 

\author{S. P. Moschou}
\affiliation{Center for Astrophysics $|$ Harvard \& Smithsonian, Cambridge, MA, 02138, USA} 

\author{C. Garraffo}
\affiliation{Center for Astrophysics $|$ Harvard \& Smithsonian, Cambridge, MA, 02138, USA} 

\author{O. Cohen}
\affiliation{Lowell Center for Space Science and Technology, University of Massachusetts, Lowell, MA 01854, USA}



\begin{abstract}
Planets in close proximity to their parent star, such as those in the habitable zones around M dwarfs, could be subject to particularly high doses of particle radiation. We have carried out test-particle simulations of $\sim$GeV protons to investigate the propagation of energetic particles accelerated by flares or travelling shock waves within the stellar wind and magnetic field of a TRAPPIST-1-like system. Turbulence was simulated with small-scale magnetostatic perturbations with an isotropic power spectrum. 
We find that only a few percent of particles injected within half a stellar radius from the stellar surface escape, and that the escaping fraction increases strongly with increasing injection radius. Escaping particles are increasingly deflected and focused by the ambient spiralling magnetic field as the superimposed turbulence amplitude is increased. In our TRAPPIST-1-like simulations, regardless of the angular region of injection, particles are strongly focused onto two caps within the fast wind regions and centered on the equatorial planetary orbital plane. Based on a scaling relation between far-UV emission and energetic protons for solar flares applied to M dwarfs, the innermost putative habitable planet, TRAPPIST-1e, is bombarded by a proton flux up to 6 orders of magnitude larger than experienced by the present-day Earth. We note two mechanisms that could strongly {\it limit} EP fluxes from active stars: EPs from flares are contained by the stellar magnetic field; and potential CMEs that might generate EPs at larger distances also fail to escape. 
\end{abstract}

\keywords{...}


\section{Introduction} \label{sec:intro}

The definition of planet habitability has been based in the last decades on the orbital distance \citep[or habitable zone, hereafter HZ,][]{Kasting.etal:93} at which the steady stellar irradiation allows for a temperature consistent with the presence of liquid water on the planetary surface. However, charged energetic particles (hereafter EPs) produced by stellar flares or shock waves driven by Coronal Mass Ejections (hereafter CMEs) and travelling into the interplanetary medium may significantly impact the conditions for life to exist in planets beyond the solar system \citep{Segura.etal:10,Ribas.etal:16,Lingam.Loeb:18a}. 

In the case of the solar wind, {\it in-situ} measurements of EP irradiation are used to assess shielding requirements for astronauts at $1$ AU \citep{Mewaldt:06,Mewaldt.etal:07}. Multi-spacecraft observations of solar eruptive events during the solar maximum of cycle $23$ ($2002 - 2006$) show that between $0.4$ and $20\%$ of the kinetic energy of CMEs in the energy range $10^{31} - 10^{32}$ erg (in the solar wind frame) is expended in accelerating solar EPs \citep{Mewaldt.etal:08,Emslie.etal:12}.

Stellar EPs are in some cases expected to cause depletion of planetary ozone layers \citep{Segura.etal:10,Tilley.etal:17}. Such depletion allows penetration of UV radiation with consequent degradation of proteins  \citep{Kerwin.Remmele:07} but also, in contrast, catalysis of pre-biotic molecules \citep{Airapetian.etal:16,Lingam.etal:18}.  \citet{Loyd.etal:18} note that ozone depletion by photolysis alone was expected to be significant only for very major flares expected to occur monthly or yearly, but note that effects of very commonly occurring weaker flares in their study could be enhanced by EPs.
Such multiple lines of evidence suggest that EPs are a component of the star/planet interaction worthy of detailed  investigation in relation to habitability. 

Propagation of EPs from the injection location to a planet is mediated by the large-scale and the turbulent components of the stellar magnetic field.  Studies of the effect of EPs on the ionization of protoplanetary disks \citep{Turner.Drake:09} or on the synthesis of short-lived nuclides in the early solar system \citep[see, e.g.,][]{Dauphas.Chaussidon:11} assumed that EPs propagate rectilinearly, unimpeded by the magnetic field structure. However, both the components of the magnetic field have been shown to lead to an efficient confinement of EPs close to young active stars \citep[see, e.g.,][]{Fraschetti.Drake.etal:18}. 

M dwarfs, the most abundant and long-lived stars in the Milky Way, are currently among the primary targets in exoplanet searches.  This is largely due to their small radius that increases the likelihood of detecting orbiting Earth-sized planets with transit techniques, or due to their low mass compared with other spectral types that increases a planet-induced radial velocity Doppler shift in the stellar spectrum. 

\cite{Youngblood.etal:17} have recently used the MUSCLES (Measurements
of the Ultraviolet Spectral Characteristics of Low-mass Exoplanetary Systems)  Treasury Survey \citep{France.etal:16} to determine that large flares on M dwarfs, i.e., with a soft $X$-ray (hereafter SXR) peak flux $\geq 10^{-3}$~W m$^{-2}$ at $1$ AU or class X10.0 in the GOES (Geostationary Operational Environmental Satellite) classification, lead to a $> 10$~MeV proton flux on planets in the HZ up to $\sim 4$ orders of magnitude higher than the present-day Earth. 

Likewise, the assumption of a solar-like correlation for T~Tauri stars between peak emission of large flares ($X$-ray luminosity $ > 10^{30}$ erg~s$^{-1}$) and energetic proton enhancements \citep{Feigelson.etal:02,Turner.Drake:09} leads to suggest an enrichment by $\sim 4$ orders of magnitude over the present-day proton density at $1$ AU.  These fluxes imply that the ionization of protoplanetary disks can locally exceed ionization due to stellar $X$-rays as a result of EPs being channeled and concentrated by magnetic turbulence   \citep{Fraschetti.Drake.etal:18}. 

Such cases show that the EPs emitted by stars more active than the Sun can play a crucial role in the evolution of the circumstellar medium, or inner ``astrosphere'' (here within $\sim 100$ stellar radii), and potentially in the habitability of exoplanets. However, while active stars might generate copious EPs, it is necessary to understand how they propagate within the stellar and interplanetary magnetic field in order to assess their potential impact.

The seven Earth-sized transiting exoplanets recently discovered  in the TRAPPIST-1 system \citep{Gillon.etal:17} are surprisingly packed within a distance of $0.062$ AU from the host star \citep{Delrez.etal:18}. Three planets (TRAPPIST-1e, f, g) have been found to orbit the HZ, that spans the range  $\sim 0.029-0.047$~AU \citep{Delrez.etal:18}, raising the question whether the enhanced EP flux at such a close distance affects the atmosphere and planetary habitability.

In this work we determine the flux of EPs impinging onto the HZ planets in the TRAPPIST-1 system by using a realistic and turbulent magnetized wind model of an M dwarf star proxy for the yet poorly-constrained wind of TRAPPIST-1. We adopt the extended magnetic field structure computed using a three-dimensional magnetohydrodynamic (MHD) model previously calibrated to the solar wind and recently applied to study the coronal structure, winds, and inner astrospheres of Sun-like stars \citep{Alvarado.etal:16a, Alvarado.etal:16b} and M-dwarfs \citep{Garraffo.etal:16,Garraffo.etal:17}, together with the propagation of EPs in stellar turbulence \citep{Fraschetti.Drake.etal:18}. We directly solve for the propagation of individual EPs in the turbulent inner astrosphere of an M dwarf wind. The turbulence is calculated via the prescription  defined in \cite{Giacalone.Jokipii:99,Fraschetti.Giacalone:12}.

In section \ref{sec:windmod}, general properties of the MHD model simulations are outlined. Section \ref{sec:environ} describes the assumptions adopted regarding EP propagation and the magnetic turbulence. Section \ref{nummeth} presents the numerical model. Section \ref{sec: results} contains the main results and \ref{sec:flux} quantifies the flux impinging on the HZ planets in the TRAPPIST-1 system. Discussion and conclusion are in Sections \ref{sec:Discussion} and \ref{sec:Conclusion}, respectively.

\begin{figure}
	\includegraphics[width=9.cm]{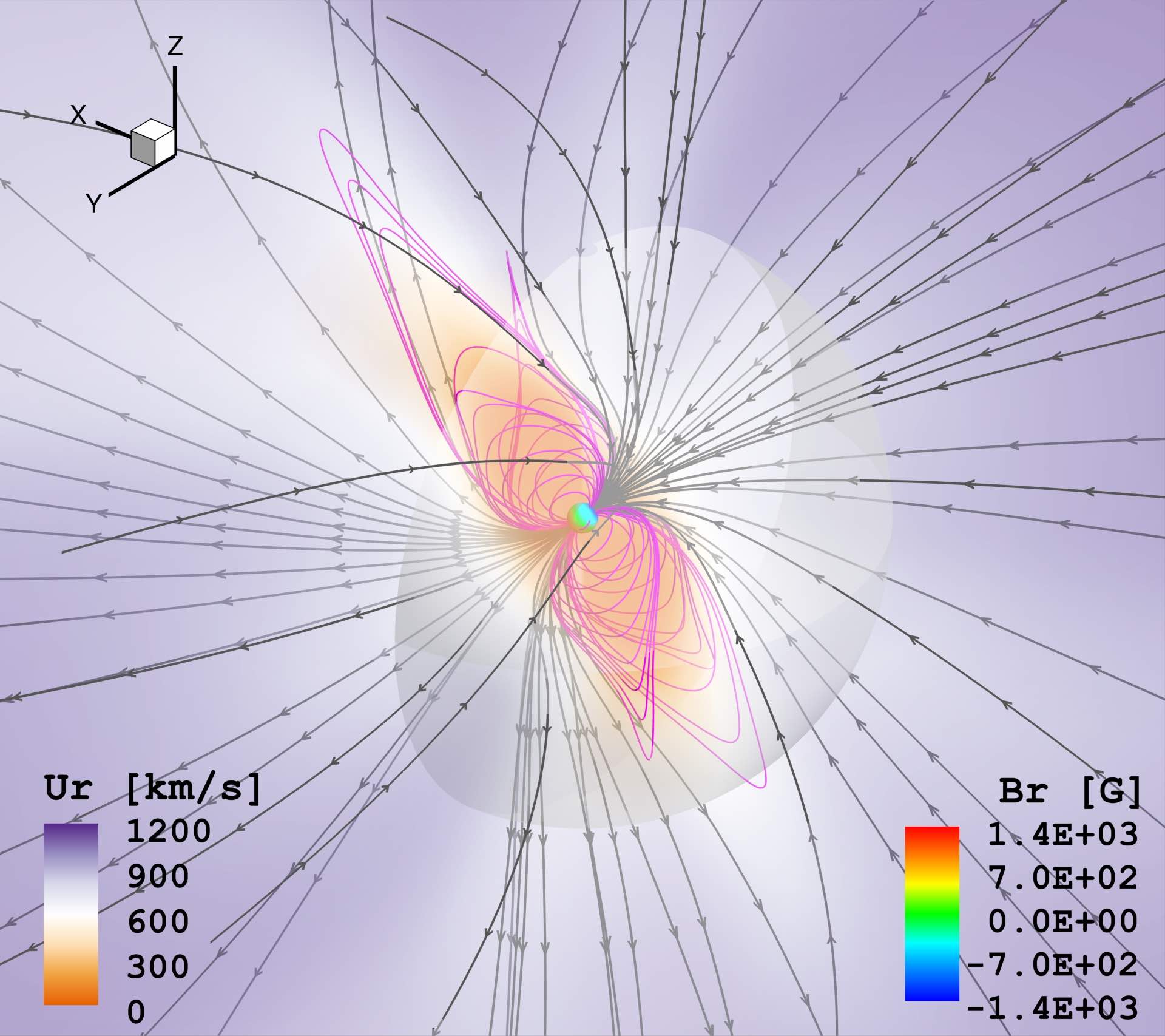}
	\includegraphics[width=9.cm]{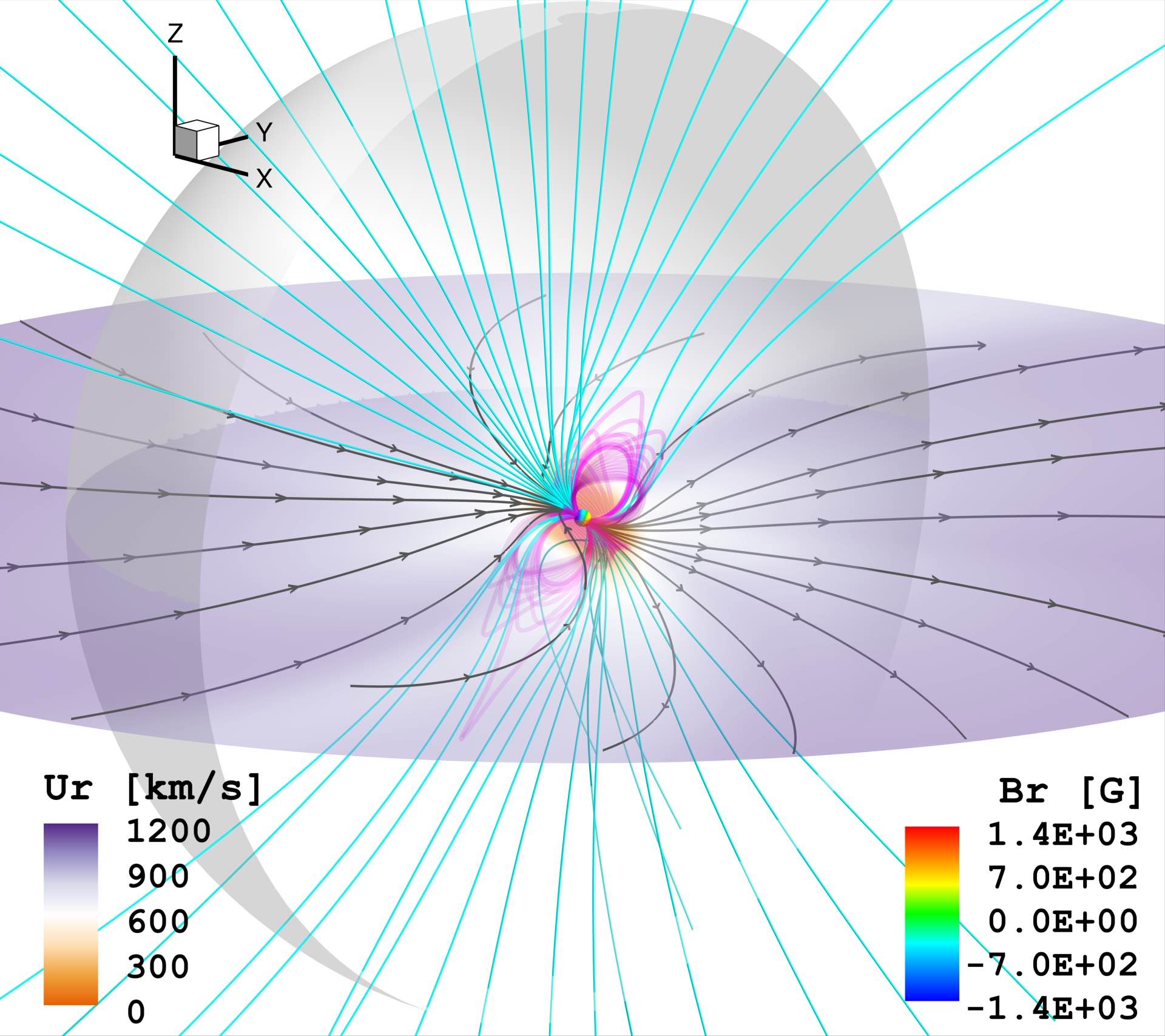}
\caption{Three dimensional stellar wind solution for GJ 3622 used here and in \citet{Garraffo.etal:17} as a proxy for TRAPPIST-1. The $Z$-axis is aligned with the stellar rotation axis. {\bf Up}: The inner sphere represents the surface of the star, color-coded by the radial component of the magnetic field ($B_{\rm r}$), at bottom-right. A slice perpendicular to the line-of-sight is included, which contains the distribution of the radial component of the wind speed ($U_{\rm r}$) as indicated by the bottom-left color-scale. The white translucent half-sphere at $R = 20 R_{*}$ denotes the maximum $R$ at which the transition between closed (magenta) and open (black with arrows) magnetic field lines is observed in the simulation. The entire field of view of the visualization is $75 R_{*}$. {\bf Bottom}: Same color code for $B_{\rm r}$ and $U_{\rm r}$ as the upper panel. The distribution of $U_{\rm r}$ is projected on the equatorial plane (plane $z = 0$). Open field lines contained in the equatorial plane are denoted by black arrows. Open field lines extending to different latitudes (cyan) are probed on the white translucent half-sphere surface $R = 60 R_{*}$ to ease visualization. Selected closed field lines are shown in magenta.  The entire field of view of the visualization is $135 R_{*}$. }
 \label{f:windmag}
\end{figure}

\section{TRAPPIST-1 MAGNETOSPHERIC MODEL}
\label{sec:windmod}

TRAPPIST-1 is a low-mass M dwarf ($0.089 M_{\odot}$) with a $3.3$~day rotation period and a radius $R_* \sim 0.114~R_{\odot}$ according to the latest observations \citep{Luger.etal:17}. It was confirmed to host seven planets orbiting in a co-planar system (within $\sim 30 $ arcmin) viewed nearly edge-on \citep{Gillon.etal:17}. All the planets reside close to the host star, with semi-major axes from 0.01~AU to 0.062~AU (Mercury orbits at 0.39~AU), with orbital periods from 1.5~days to 20~days.

As a background medium for studying the propagation of EPs within the TRAPPIST-1 system, we adopt the wind and magnetosphere model computed by \citet{Garraffo.etal:17} using the 3D MHD code 
{\sc Block Adaptive Tree Solarwind Roe Upwind Scheme} \citep[BATS-R-US, ][]{Powell.etal:99, Toth:12}, in the version that incorporates the Alfv\'en Wave Solar Model (AWSoM) \citep{vanderHolst:14}. A data-driven global MHD method is used that was initially developed to reconstruct the solar atmosphere and the solar wind. 
BATS-R-US employs a radial field magnetogram as a boundary condition for the stellar photospheric magnetic field.  In the case of application to the Sun, this is a solar magnetogram but stellar magnetograms obtained using the Zeeman-Doppler Imaging technique \citep{Donati.Brown:97} can also be used. 

Zeeman-Doppler Imaging is presently limited to luminous, fairly rapidly rotating stars. TRAPPIST-1, despite its relatively fast spin, is optically faint \citep[$M_v=18.8$, ][]{Gillon.etal:17} and out of reach of current Zeeman-Doppler Imaging capabilities.  Unfortunately both the distribution of the magnetic field on its surface and the direction of the rotation axis are unknown due to the extreme faintness of the star; moreover, both are subject to change in time with time scales of years to greater, due to the periodic change of magnetic polarity and to the axis precession, respectively. Its average magnetic field, however, has been estimated to be $\sim 600$~G using Zeeman broadening \citep{Reiners.Basri:10}. There is growing agreement that the geometry of the magnetic field depends on the rotation period and spectral type of the star \citep{Vidotto.etal:14,Garraffo.etal:15,Reville.etal:15, Finley.Matt:18}.  
\citet{Garraffo.etal:17} therefore used as a proxy for TRAPPIST-1 the magnetogram observed for GJ~3622 \citep{Morin.etal:10}, an M4 dwarf with a rotation period of $1.5$~days. The field on its surface reaches a maximum of $1.4$~kG, yielding an average field of $\sim$600~G, consistent with the TRAPPIST-1 observations. The magnetic structure is not expected to change significantly between stars with periods of 1 to 3 days.  We note that our approach is different to that of \cite{Dong.etal:18}, who estimated the ion escape rate in the seven planets using a wind model based on a solar magnetogram under solar minimum conditions, rescaled to a magnetic field strength more like typical M-dwarf values \citep{Morin.etal:10}. 

The GJ~3622 magnetic field is vaguely dipolar with a notable misalignment between the rotation axis and the magnetic field amounting to a few tens of degrees ($\sim 40^\circ$ - $50^\circ$). The wind and magnetosphere model are illustrated in Figure~\ref{f:windmag}. 

\section{Stellar energetic particles in the TRAPPIST-1 environment} \label{sec:environ}
\subsection{General assumptions on EPs: origin and propagation}\label{assumptions}

Our general goal here is to explore the effect of small-scale magnetic turbulence on the propagation of EPs through the magnetosphere of the host star TRAPPIST-1, and as far as the outermost planet located at a distance of $\sim 0.062$~AU. In particular, we focus on a comparison of the EP flux generated at the star itself with that which propagates out to planets 1b, 1e and 1h.

Two processes are assumed to produce the non-thermal particles \citep{Fraschetti.Drake.etal:18}: 1) shock waves driven by CMEs, travelling in the interplanetary medium and therein accelerating and releasing EPs; 2) flares occurring within the stellar corona and releasing EPs within a small distance from the stellar surface ($\sim 0.5 R_*$). Both such processes are assumed to produce the $\sim$~GeV kinetic energy  protons studied here. This assumption can be justified by a solar analogy: former GOES measurements correlating solar proton enhancements at $1$ AU with SXR flares do not unequivocally pinpoint the flares as the only sources of particle acceleration as CME-driven shocks are consistent with such a correlation as well \citep{Belov.etal:07}. 

In our simulations only the location of injection of EPs (at a distance $R_s$ from the star), rather than the acceleration mechanism, is assigned. As for the abundance of accelerated particles in the circumstellar medium at a given distance from the host star, we use the estimate based on solar scaling relations between EP fluence and far-UV and SXR fluence during flares by \citet{Youngblood.etal:17}. This scaling provides a time-averaged EP enrichment for time scales comparable with a statistically typical flare duration \citep{Vida.etal:17} .

We calculate the propagation of the EPs using a test particle approach within a realistic representation of the interplanetary medium that includes magnetic field fluctuations. 
The large-scale structure used here for the  TRAPPIST-1 magnetic field (see Fig.~\ref{f:windmag}) has an approximately dipolar structure with no significant field lines wrapping around the star as might be expected for  T~Tauri stars and some fast rotators \citep[see, e.g., ][]{Gregory.etal:09,Cohen.etal:10b,Fraschetti.Drake.etal:18}. Nevertheless, it is still uncertain whether the average $\sim$~kiloGauss magnetic field of TRAPPIST-1 allows for CME escape and the outward driving of EPs accelerated at shocks \citep{Drake.etal:16,Osten.Wolk:15}. Under the assumption that EPs can be steadily supplied by flares and CMEs, the dominant magnetic effects we are concerned with for EP propagation in TRAPPIST-1 are expected to be scattering and perpendicular diffusion in the turbulent stellar field. 

The MHD wind solution and the magnetic turbulence are stationary on the time-scale of EP propagation to a good approximation. The EPs travel at speed $\simeq c$, whereas the stellar rotation speed close to the surface is $\sim 2$~km~s$^{-1}$, and the Alfv\'en wave speed in the circumstellar medium is $\sim 10^4$ km~s$^{-1}$ ($\sim 10^3$ km~s$^{-1}$) at a distance $\sim 10 R_*$ ($110 R_*$, semi-major axis of the outermost planet) from the host star. This holds for M dwarfs in general. The visible light periodograms of M dwarfs (with radii in the range $0.08 - 0.6 R_\odot$)---presumably dominated by rotational modulation signatures---typically peak at a few days over a range of periods $\sim 1$--100 days \citep{Hawley.etal:14}, with a corresponding surface rotation speed over a range $0.04 - 30$ km/s \citep{Barnes.etal:14,Jeffers.etal:18}. Only the earliest M dwarfs ($0.6 R_\odot$) with rotation periods $\leq 3$ days have surface rotation speeds $> 10$ km~s$^{-1}$. Dynamical timescales are therefore much longer than the EP travel time in our simulations (typically $<1$ hour).

\subsection{Turbulent stellar magnetic field}\label{propagation}

\begin{figure}
	\includegraphics[width=9.cm]{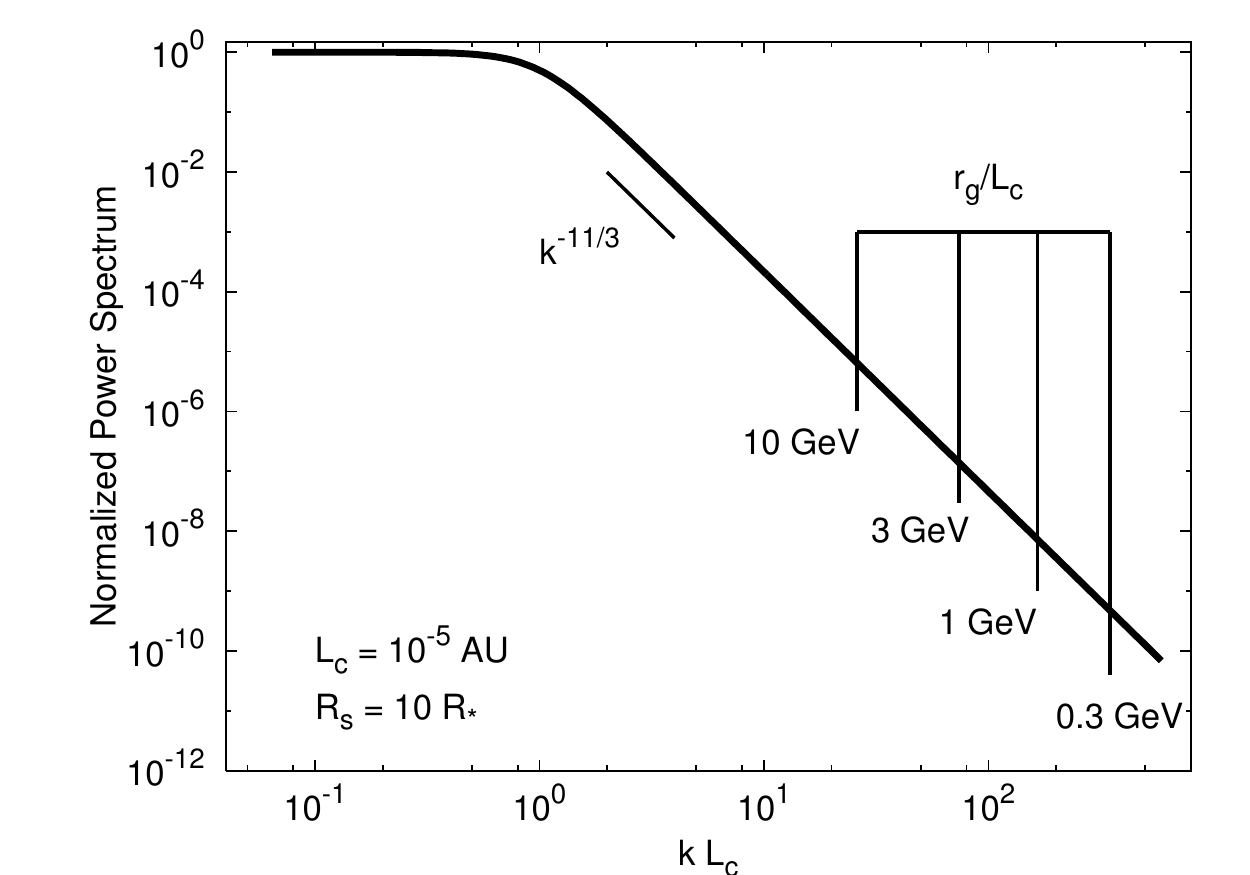}
\caption{Three-dimensional power-spectrum of the magnetic turbulence used in the test-particle simulations with Kolmogorov power-law index ($11/3$) in the inertial range (see Sect. \ref{nummeth}). The vertical lines mark the resonant wavenumbers in the average magnetic field at $R_s = 10 R_{\star}$ ($B_0 \simeq 2.2$~G) for individual protons with kinetic energies $E_k=0.3, 1, 3, 10$~GeV (here $L_c = 10^{-5}$~AU). 
 \label{Pspectrum}
 }
\end{figure}

In analogy with the measurements of interplanetary magnetic turbulence \citep[e.g., ][]{Jokipii.Coleman:68}, and of interstellar density turbulence  \citep{Armstrong.etal:95}, we prescribe a magnetic turbulence power spectrum having the shape of a power-law (Kolmogorov) in the 3D turbulent wavenumber ${\rm k}$ (see Fig.~\ref{Pspectrum}). Scale-dependent anisotropic turbulence ({\it \'a la} the \citealt[e.g.][]{Goldreich.Sridhar:95} model) explaining the origin of the solar wind MHD-scale turbulence anisotropy \citep[e.g., ][]{Horbury.etal:08}, has unsettled theoretical transport properties \citep{Laitinen.etal:13,Fraschetti:16a,Fraschetti:16b} and would require a more cumbersome numerical code.

The test-particle simulations presented here track naturally the pitch-angle scattering and cross-field motion of EPs caused by the small-scale turbulence. An alternative approach to EP transport involves Monte Carlo simulations that reproduce the pitch-angle scattering and neglect perpendicular transport \citep[see, e.g., ][]{Ellison.etal:81}. The nearly radial spread-out of the open magnetic field lines of the astrosphere used here leads to an observable consequence (see Sect. \ref{sec: results}) of the turbulent transport across field lines \citep{Fraschetti.Jokipii:11,Strauss.etal:17}. In contrast, in the case of the T~Tauri star studied in \citet{Fraschetti.Drake.etal:18} the wrapping of magnetic field lines around the star prevented an assessment of the effect of the transport across field lines.

Due to the lack of observational estimates of the correlation length, or injection scale, $L_c$, of the magnetic turbulence within the circumstellar medium (see Fig. \ref{Pspectrum}), we adopt the uniform value $L_c =10^{-5}$~AU throughout the simulation box. A simulation set carried out with a smaller uniform $L_c =10^{-6}$~AU shows that the statistical properties of EPs are not significantly affected by the choice of $L_c$, provided that the resonance condition is satisfied. In this regard, $L_c =10^{-5}$ AU is a reasonable value for the quite small range in radial distance of the planets in the TRAPPIST-1 system, within $0.062$~AU. The chosen value of $L_c$ ensures resonance with turbulent inertial scales at each EP energy considered (see Fig.~\ref{Pspectrum}) during their entire propagation. Such a condition reads 
\begin{equation}
{\rm k} r_g({\mathbf x})/2\pi = r_g({\mathbf x})/L_c < 1 \, ,
\end{equation}
for each wave-number ${\rm k}$ within the inertial range; here, $r_g ({\mathbf x}) = p_\perp c/ e B_0 ({\mathbf x})$ is the gyroradius of a proton with momentum  $p_\perp$ perpendicular to the unperturbed and space-dependent magnetic field $B_0 ({\mathbf x})$ of TRAPPIST-1, $e$ the proton electric charge and $c$ the speed of light in vacuum.

The power of the magnetic fluctuation $\delta B (\vect{x})$ relative to $B_0 (\vect{x})$ is defined as 
\begin{equation}
\sigma^2 = (\delta B ({\mathbf x}) /B_0 ({\mathbf x}))^2 .
\label{sigma}
\end{equation}
Here, $\sigma^2$ is assumed to be independent of space throughout the simulation box as well. The spherical average of the unperturbed field $\langle B_0 ({\bf x}) \rangle_{\Omega}$ produced by the 3D-MHD simulations (see Sect.~\ref{sec:windmod}) drops with radius $R$ from $2 R_*$ as $\sim R^{-2.2}$. On the other hand, the solar wind measurements yield for the turbulence amplitude $\delta B$ between $0.3$ and $4$ AU a power-law dependence on heliocentric distance with a very similar index ($\simeq −2.2$) at a variety of helio-latitudes \citep{Horbury.Tsurutani:01}. Thus, in the lack of any current measurement of the magnetic turbulence around  TRAPPIST-1, it seems reasonable to assume a uniform $\sigma^2$, following \citet{Fraschetti.Drake.etal:18}. 

The turbulence might be generated by the stirring of the plasma at the outer scale $L_c$, followed by a cascade, or by plasma instabilities at kinetic scales generated, e.g., by streaming of EPs along the field; we neglect the latter here as we are restricted to the test-particle limit. The turbulence within the violently active M dwarf magnetosphere is likely to be much
stronger than that in the solar wind \citep[$\sigma^2$ not greater than $0.1$, ][]{Burlaga.Turner:76}. Thus, we considered values of $\sigma^2$ spanning the range $0.01 - 1.0$. The interpretation of our simulations makes use of the  scattering mean free path, $\lambda_\parallel$, given by quasi-linear theory \citep{Jokipii:66}, that reads \citep{Giacalone.Jokipii:99,Fraschetti.Drake.etal:18}
\begin{equation}
\lambda_\parallel ({\mathbf x}) \simeq 4.8 (r_g({\mathbf x})/L_c)^{1/3} L_c/\sigma^2 \, .
\label{lambda}
\end{equation}
The choices of uniform $L_c$ and $\sigma^2$ imply that $\lambda_\parallel$ depends on spatial coordinates only via $r_g (\vect{x})$ (i.e., $B_0 (\vect{x})$).

\begin{figure}
\hspace{-3cm}	\includegraphics[width=16.cm]{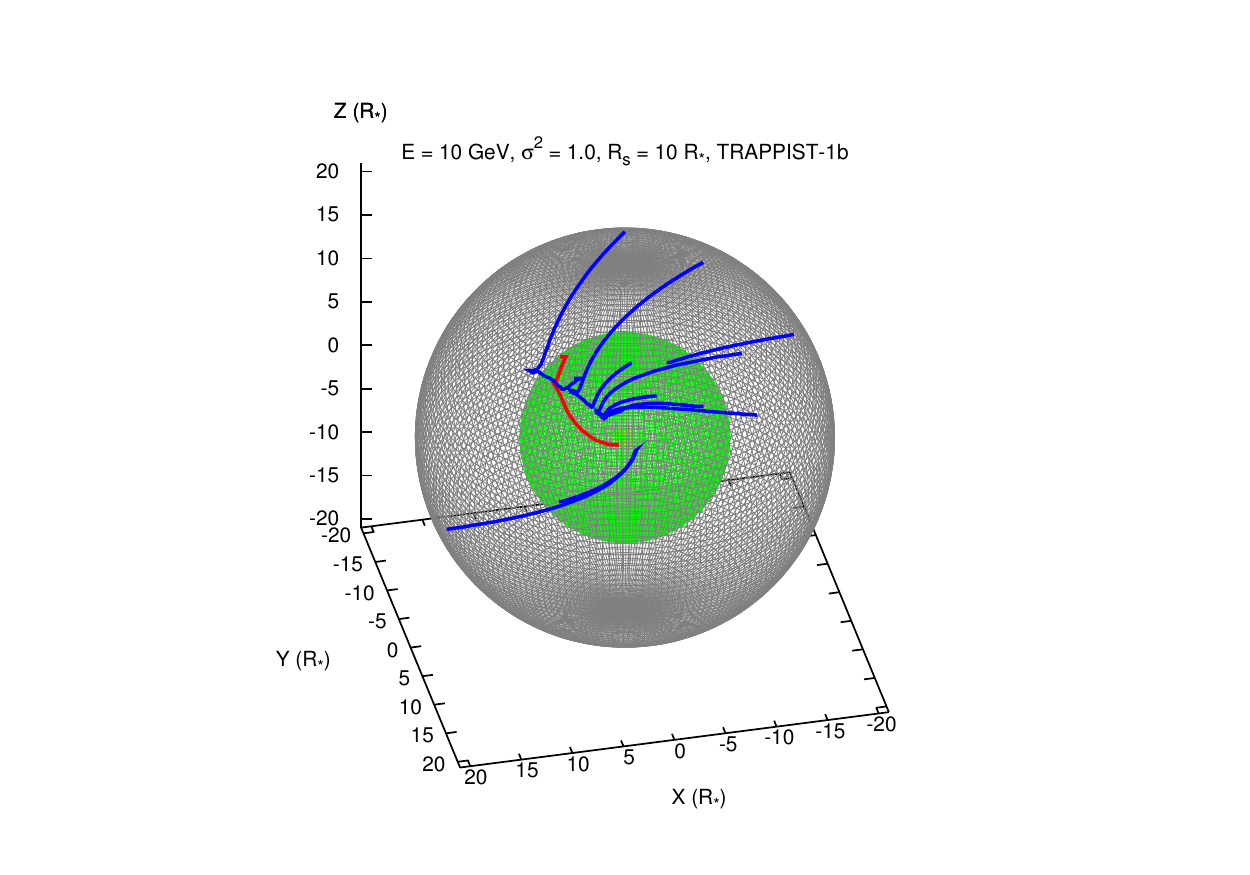}
\caption{Three-dimensional trajectories of selected 10~GeV kinetic energy protons injected at $R_s = 10 R_{\star}$ (green sphere) and hitting (in blue) the sphere at $R_p = R_b = 20 R_{\star} = 0.011 $ AU (in gray); here $\sigma^2 = 1.0$. We plot in red the trajectory of EPs collapsing back onto the star.
 \label{trajectory_1d4_b}
 }
\end{figure}

\begin{figure}
\hspace{-3cm}	\includegraphics[width=16.cm]{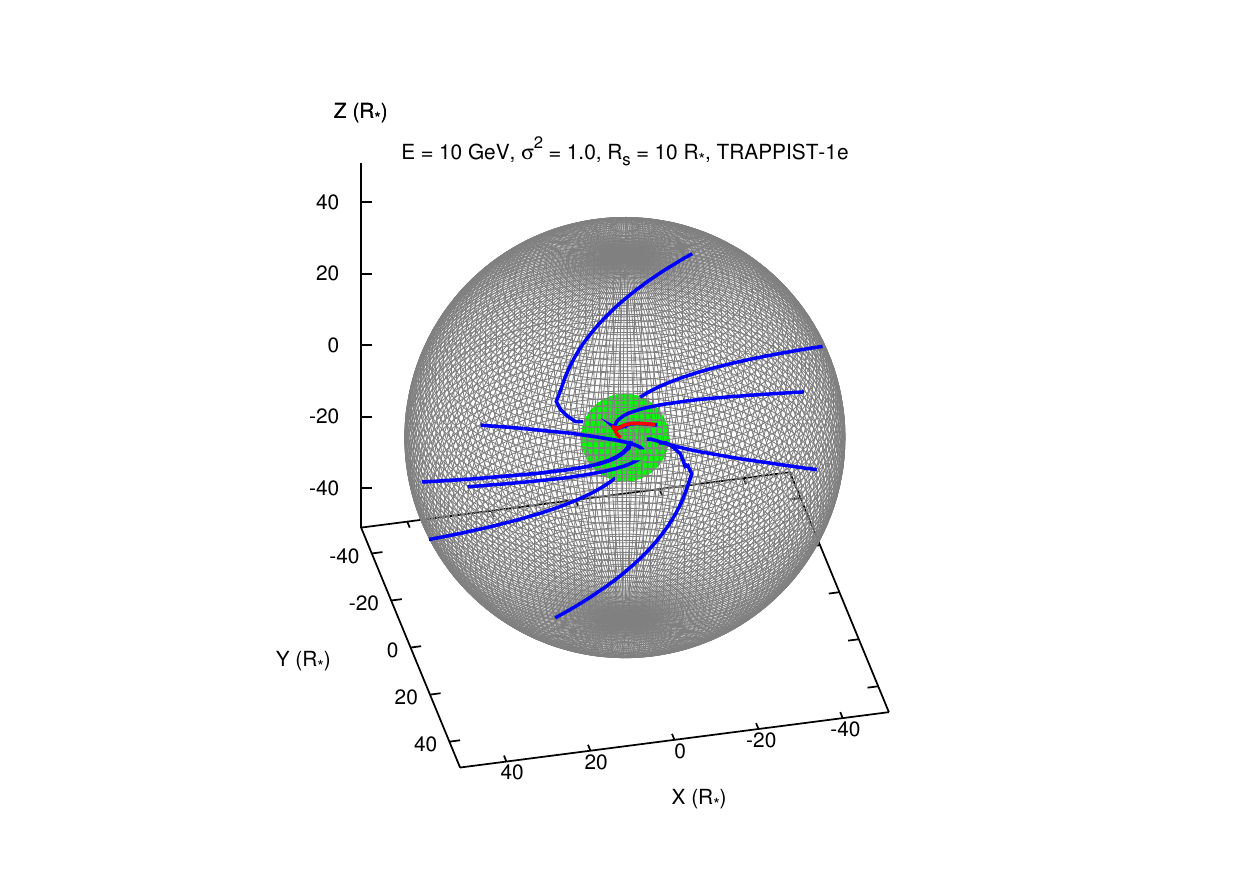}
\caption{Same as Fig. \ref{trajectory_1d4_b} for $R_p = R_e = 51 R_{\star} = 0.029$ AU. 
 \label{trajectory_1d4_e}
 }
\end{figure}

\begin{figure}
\hspace{-3cm}	\includegraphics[width=16.cm]{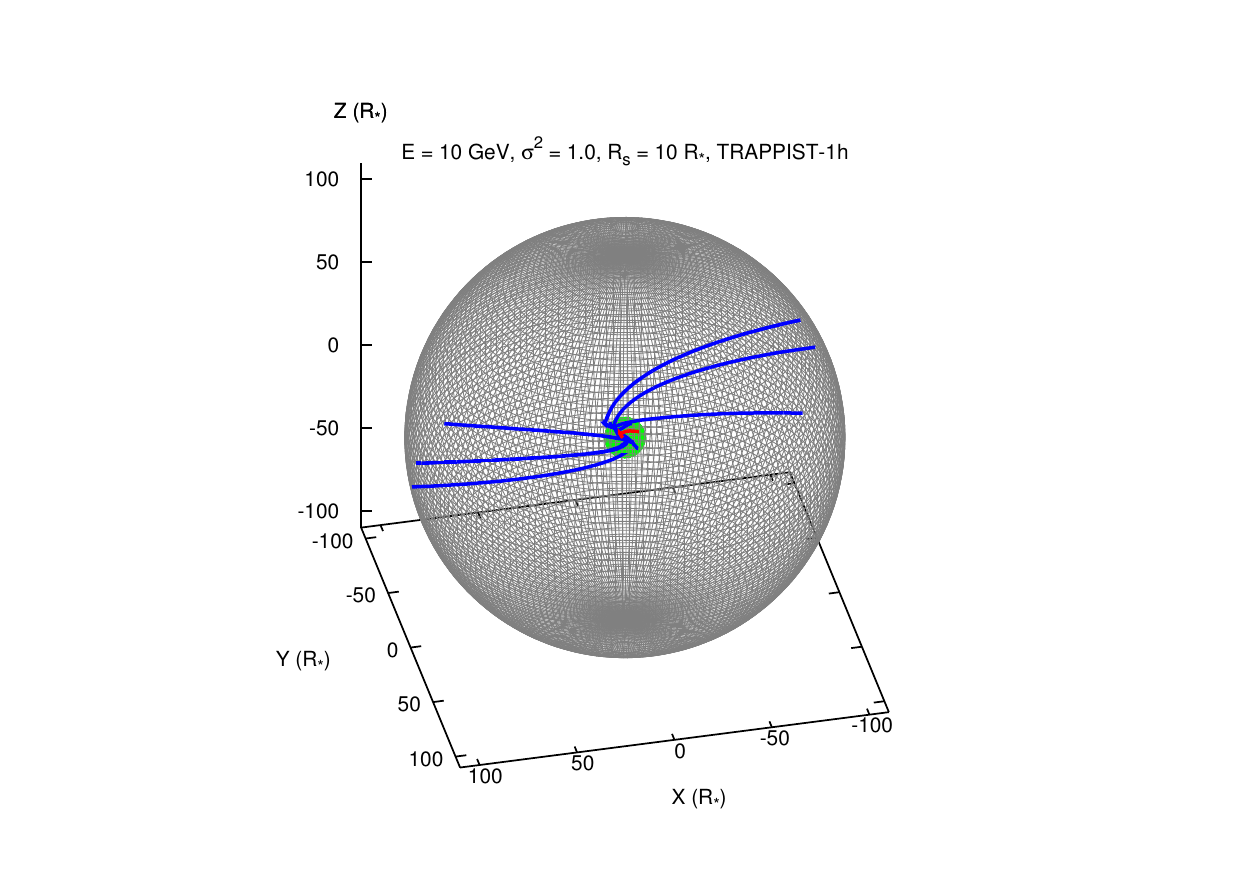}
\caption{Same as Fig. \ref{trajectory_1d4_b} for $R_p = R_h = 110 R_{\star} = 0.062$ AU. 
 \label{trajectory_1d4_h}
 }
\end{figure}

\section{Numerical Method}\label{nummeth}

In our numerical experiments, we have directly integrated the trajectories of $\sim 10^4$ energetic protons propagating in a turbulent magnetic field that can be decomposed as  
\begin{equation}
{\bf B(x) = B}_0 ({\bf x}) + \delta {\bf B(x)},
\end{equation}
where the large-scale component, ${\bf B}_0 ({\bf x})$, is the 3D magnetic field generated by the 3D-MHD simulations as calculated in \citet{Garraffo.etal:17} and described in Section~\ref{sec:windmod}; the random component ${\bf \delta B} = {\bf \delta B} (x, y, z)$ has a zero mean ($\langle \delta {\bf B(x)} \rangle = 0$). Here ${\bf \delta B} (x, y, z)$ is calculated as the sum of plane waves with random orientation, polarization, and phase following the prescription in \citet{Giacalone.Jokipii:99,Fraschetti.Giacalone:12}. We use an inertial range $k_{\rm min} < k < k_{\rm max}$, with $k_{\rm max}/k_{\rm min} = 10^2$, where $k_{\rm min} = 2\pi/L_c$ and $k_{\rm max}$ is the magnitude of the wavenumber corresponding to some turbulence dissipation scale. In \cite{Fraschetti.Drake.etal:18} we verified that an inertial range extended by one decade to smaller scales does not substantially change the resulting distribution of a large number of EPs hitting a protoplanetary disk, despite being computationally much more expensive; we assume that a larger inertial range is not relevant for the M dwarf circumstellar turbulence either. 

The turbulence power spectrum within the inertial range (Fig.~\ref{Pspectrum}) is assumed to be a three-dimensional Kolmogorov power-law (index $-11/3$). At scales larger than $k_{\rm min} ^{-1}$ ($k_{0} < k < k_{\rm min}$), the power spectrum is taken as constant (see, e.g., \citet{Jokipii.Coleman:68} for the solar wind case).

In our simulations, the EPs are injected uniformly on spherical surfaces at a variety of radii, $R_s$, with a velocity distribution isotropic in pitch-angle. The number of particles is then rescaled by using the enhancement in EP flux inferred at a given distance from the star in \citet{Youngblood.etal:17}. After propagation through the inner astrosphere, the EP angular location is recorded on spherical surfaces at distances $R_p$. We verified that the particle energy is conserved to a relative accuracy of ~$10^{-3} - 10^{-4}$.

\section{Results}\label{sec: results}

\subsection{Particle Trajectories}

Figures \ref{trajectory_1d4_b} to \ref{trajectory_1d4_h} show the trajectories of selected individual EPs injected at $R_s = 10 R_{\star} = 0.0056$~AU with $\sigma^2 = 1.0$. All EPs are allowed two possible fates in our simulations: hitting (in blue) the spherical surface at $R_p = R_{b,e,h}$, where $R_{b,e,h}$ equals the semi-major axes of the planets TRAPPIST-1b, e, h (respectively $20.4 R_* = 0.0115$ AU, $51.8 R_* = 0.02916$ AU  and $110. R_* = 0.0617$ AU \citep{Delrez.etal:18}), or  collapsing (in red) back to the star.

\subsection{Weak Turbulence Case}\label{sec:weak}

Figures ~\ref{2D_E_1d3}, \ref{2D_E_1d4} and \ref{2D_E_1d4_5} show the spherical coordinates of the hitting points for 1~GeV (Fig.~\ref{2D_E_1d3}) and 10~GeV (Figs.~\ref{2D_E_1d4} and \ref{2D_E_1d4_5}) kinetic energy protons, injected at $R_s = 10 R_{\star}$ (Figs.~\ref{2D_E_1d3}, \ref{2D_E_1d4}) and $R_s = 5 R_{\star}$ (Fig.~\ref{2D_E_1d4_5}) recorded at the spheres $R_p = R_{b,e}$. The total number of injected EPs ($N_{\rm inj}$) is the same in all cases. Different rows correspond to different values of $\sigma^2$, increasing from top to bottom; different columns correspond to a different planet, 1b (left) and 1e (right). The colorbar is scaled to the maximum number of EPs per pixel, and varies strongly between panels; thus, the same color in different panels does not indicate the same absolute number of EPs. 
The plane $\theta' = \theta + 90^\circ = 90^\circ$ perpendicular to the stellar rotation axis, where $-90^\circ < \theta < 90^\circ $ is the latitude, marks the plane of the planetary orbits within $30$~arcmin \citep{Delrez.etal:18}. 

In Figs. \ref{2D_E_1d3} and \ref{2D_E_1d4},  
for weak turbulence ($\sigma^2 = 0.01$, upper row), the distribution of hitting points spreads fairly uniformly over the  $R_p$-sphere. Such a distribution mirrors the uniform distribution of the injection points of EPs and results from the EPs propagation outward close to the scatter-free limit, i.e., uniform and static electric and magnetic field, along the open field lines intercepted on the sphere at $R_s$ (greater $\lambda_\parallel$ for small $\sigma^2$, from Eq.~\ref{lambda}).

The perpendicular diffusion coefficient $\kappa_\perp$ grows, regardless of the model, as $\kappa_\perp \sim \sigma ^2$ \citep{Giacalone.Jokipii:99,Fraschetti.Jokipii:11,Strauss.etal:17} leading to a negligible decorrelation of EPs, for small $\sigma^2$ from the direction of the average magnetic field. Thus, the resulting distribution of hitting points at $R_p$ is close to the injection distribution at $R_s$ and the trajectories nearly map the unperturbed magnetic field ${\vect B}_0$. However, we note that the ratio of the number of EPs at $R_p$-sphere ($N_{R_p}$) to $N_{\rm inj}$ is limited to $20-25\%$ (see also Fig. \ref{Fraction_E_1d4}, left panel), as a large fraction ($75-80\%$) collapse back to the star. The latter EPs are released on closed field lines that are prevalent at $R_s = 10 R_*$ (see Fig. \ref{f:windmag}), and propagate along those closed field lines back to the star, due to the large $\lambda_\parallel$ (see Eq. \ref{lambda}) and negligible perpendicular diffusion. 

We also note in Fig. \ref{Fraction_E_1d4}, left panel, that for each value of $\sigma^2$ the ratio $N_{R_p}/N_{\rm inj}$ decreases for greater $R_p$, i.e., decreasing from 1b (red) to 1h (blue). This occurs because some EPs that propagate past an inner $R_p$-sphere undergo pitch-angle diffusion that leads them to move backward and to collapse to the star without reaching the outer $R_p$-sphere. In addition, Fig.~\ref{Fraction_E_1d4}, left panel, shows a smaller difference for each value of $\sigma^2$ between the blue and green curves as compared with green and red ones: this change results from the transition of the large-scale $B_0$-field structure from closed/open to prevalently open field lines between the 1b (red) and 1e (green), whereas between 1e and 1h (blue) all field lines are open (cfr. Fig. \ref{f:windmag}), so that no significant difference is expected between the green and the blue curves. We note that the likelihood of backward trajectories decreases further out due to the increase of the mean free path: $\lambda_\parallel$ increases outward as $r_g^{1/3} \propto B_0^{-1/3}$ (see Eq. \ref{lambda}) for $B_0$ decreasing outward in a uniform $\sigma^2$, so most EPs channelled onto an open line that reach 1e will also reach 1h.

We have run an additional set of simulations with $R_s = 1.5 R_*$, i.e., at a distance of $0.5 R_*$ from the stellar surface, for particles with $E = 0.3$~GeV.  For these simulations, negligible turbulence was adopted ($\sigma^2 = 10^{-8}$) since within the chosen turbulent inertial range the EPs would not scatter resonantly as  $r_g$ is suppressed by the strong  $B_0$ field close to the surface. We find that the ratio $N_{R_p}/N_{\rm inj}$ is in the range $3.0 - 3.7\%$ for $R_p = R_b$ or $R_h$.

\subsection{Effect of Stronger Turbulence}

\begin{figure*}
	\includegraphics[width=9.1cm]{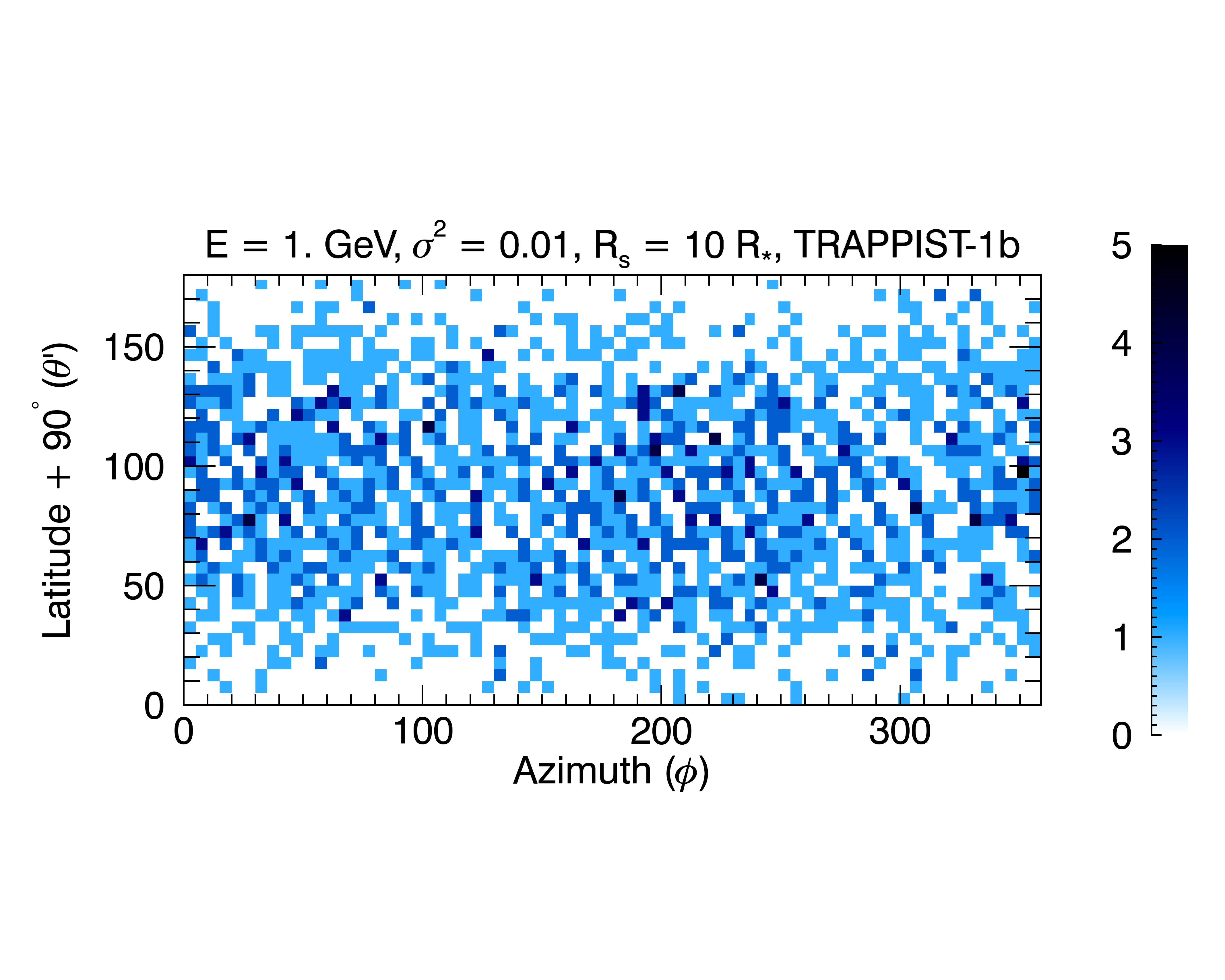}
	\includegraphics[width=9.1cm]{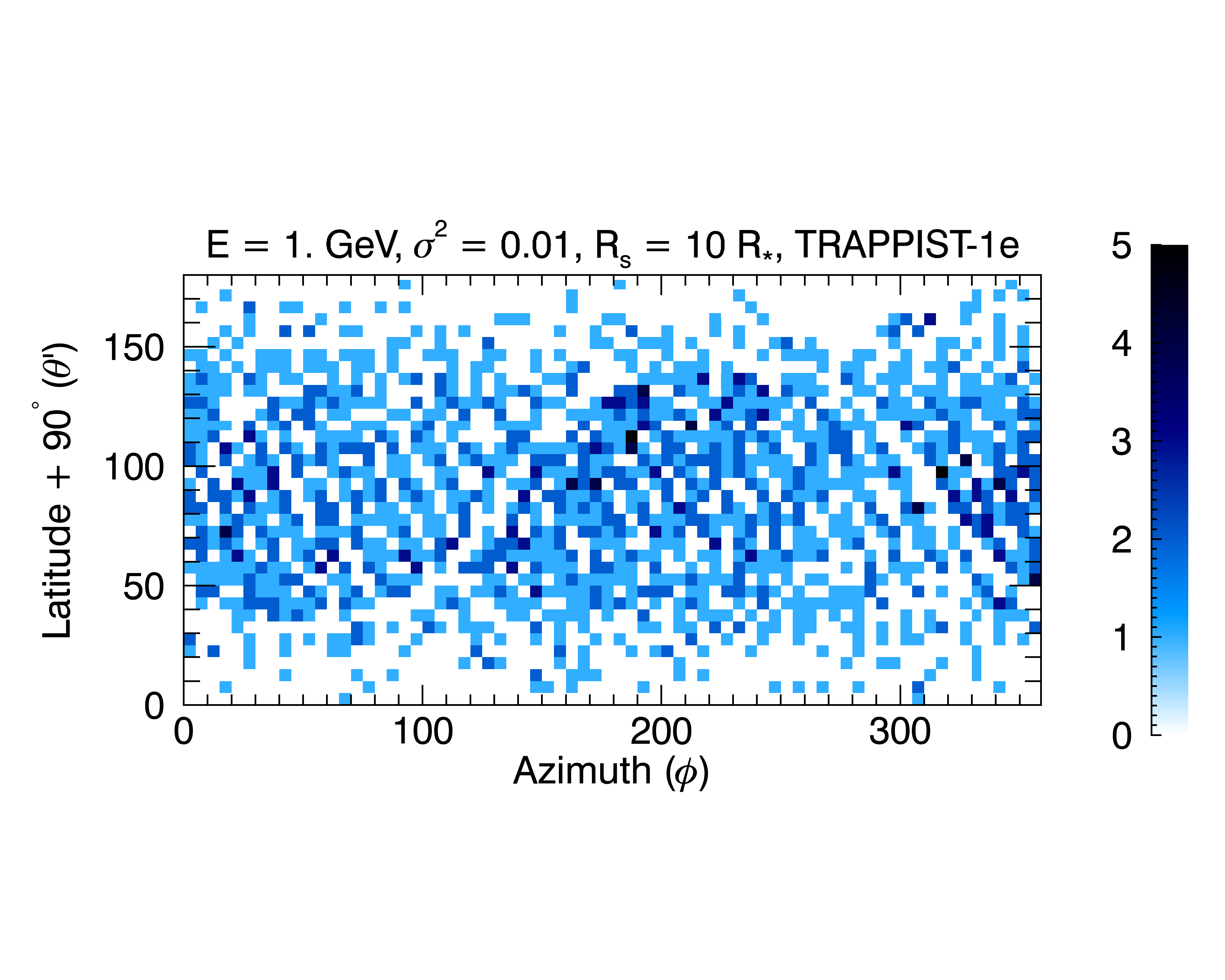}\\
	\includegraphics[width=9.1cm]{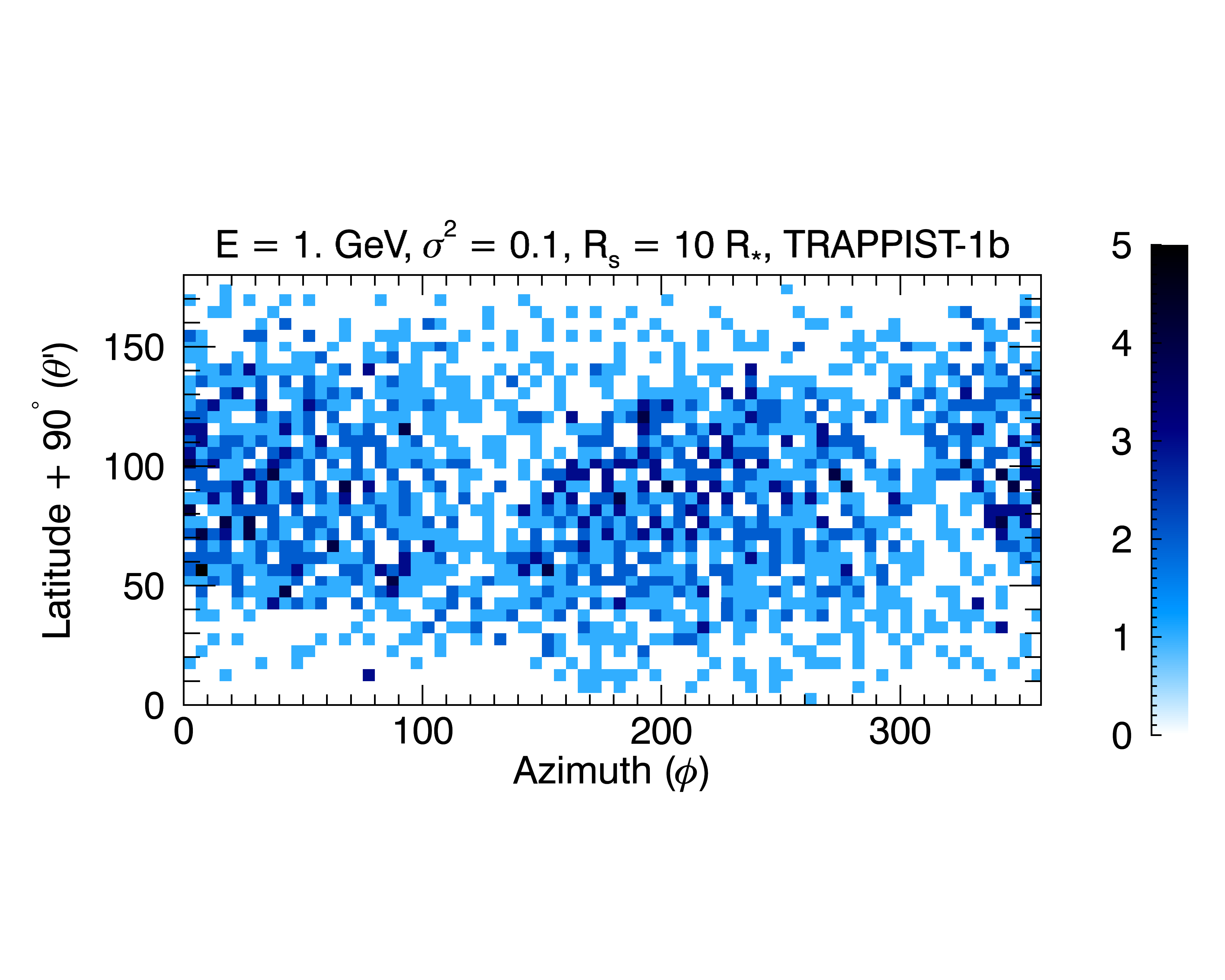}
	\includegraphics[width=9.1cm]{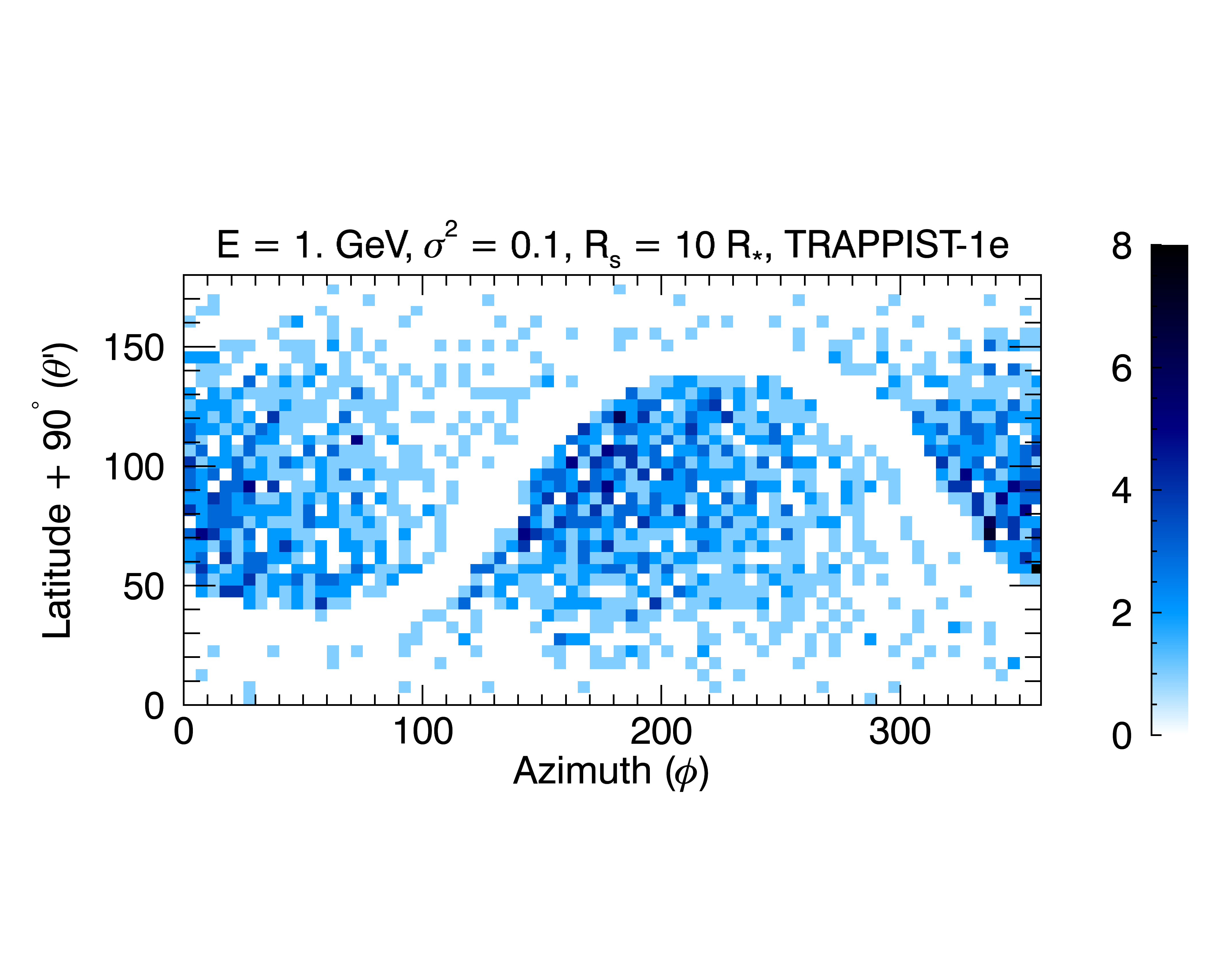}\\
	\includegraphics[width=9.1cm]{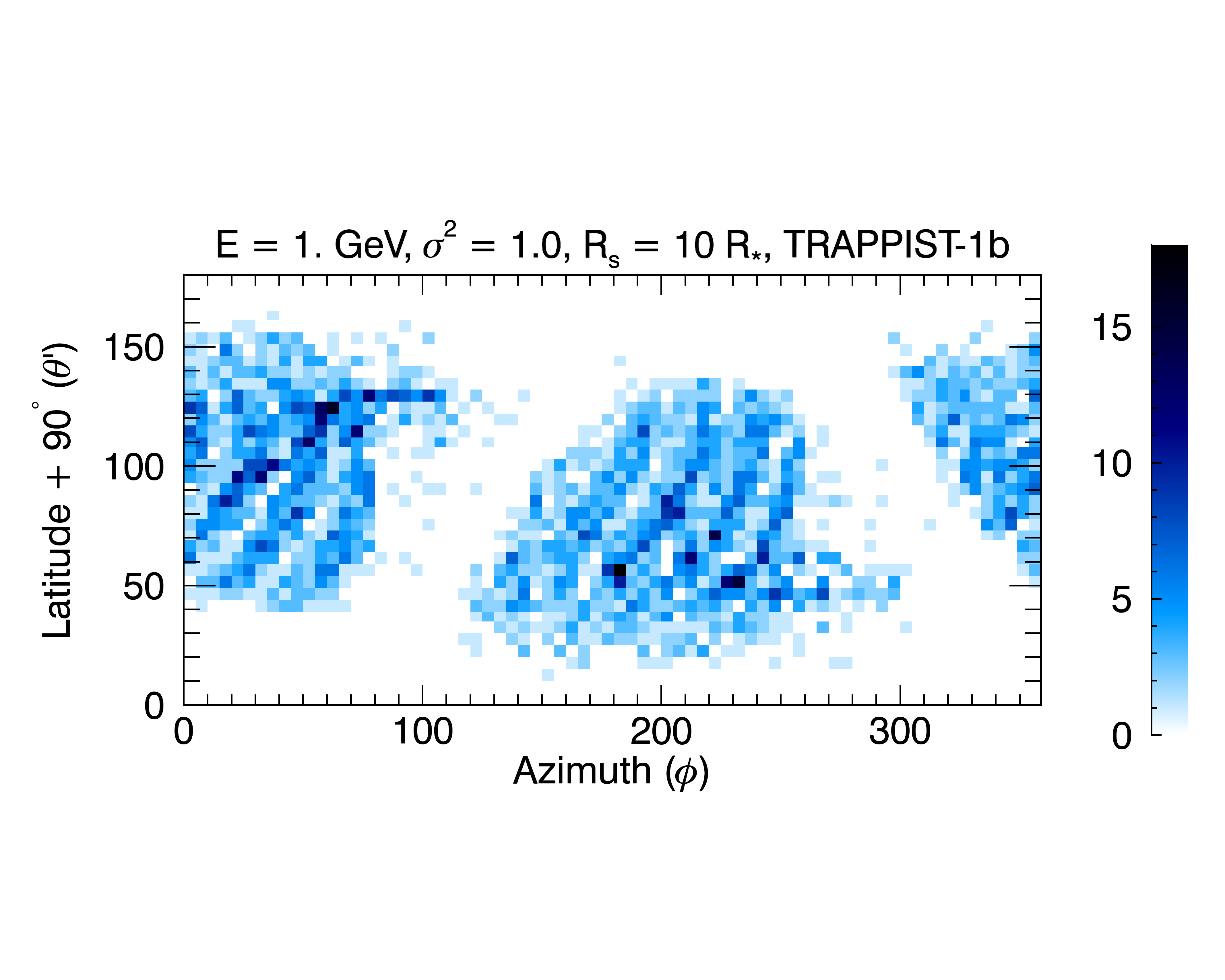}
	\includegraphics[width=9.1cm]{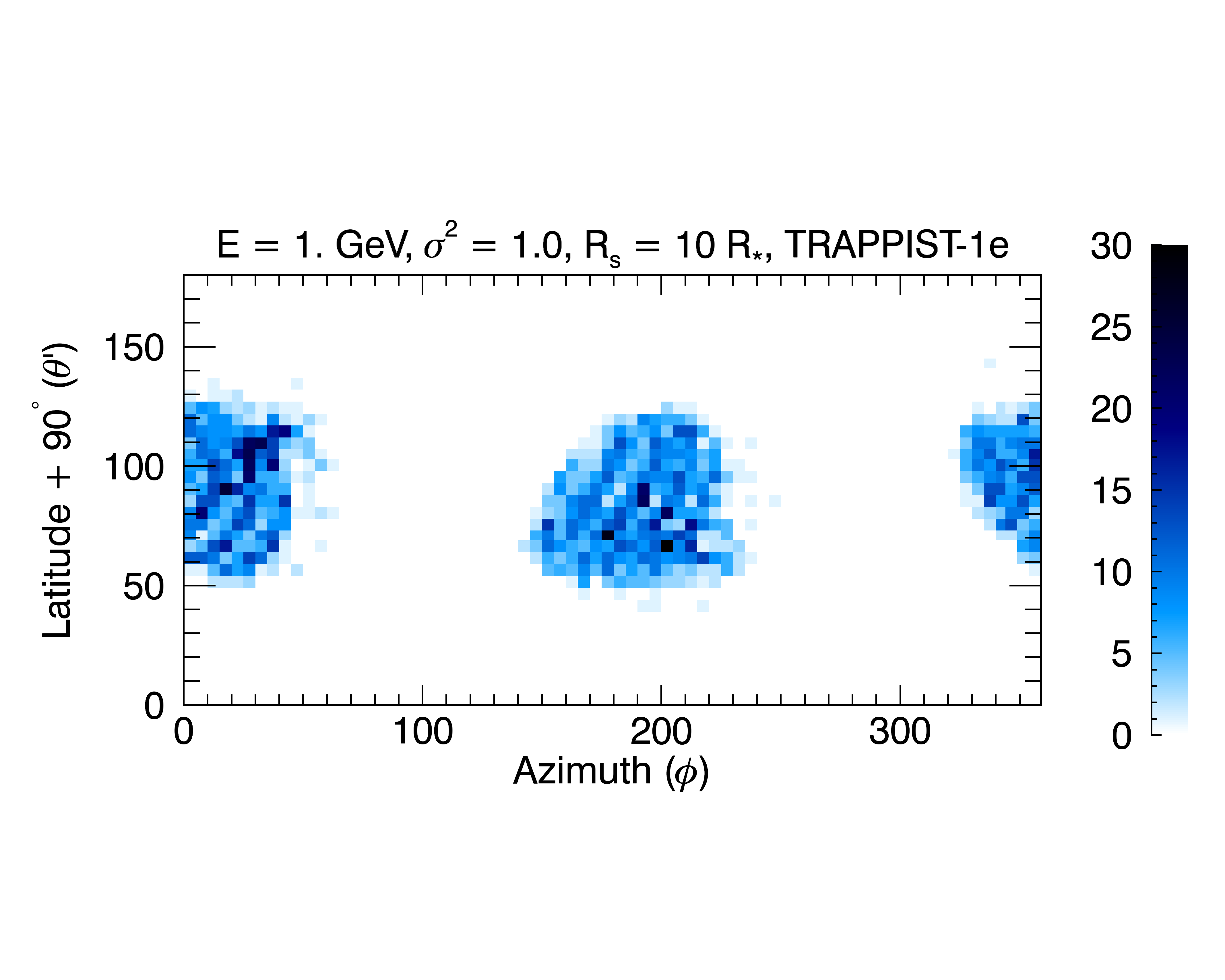}
\caption{Coordinates of the hitting points for 1~GeV kinetic energy protons, injected at $R_s = 10 R_{\star}$ with $L_c = 10^{-5}$~AU, at the spherical surface with radius $R_p = R_b$ (left column) and $R_p = R_e$ (right column) and for various values of $\sigma^2 $: $\sigma^2 = 0.01$ (upper row), $\sigma^2 = 0.1$ (middle row) and $\sigma^2 = 1$ (lower row). The $x$ ($y$) axis indicates the azimuthal (polar) coordinates on that sphere. The colorbar measures the number of EPs relative to the maximum in each panel.  \label{2D_E_1d3}}
\end{figure*}

\begin{figure*}
	\includegraphics[width=9.1cm]{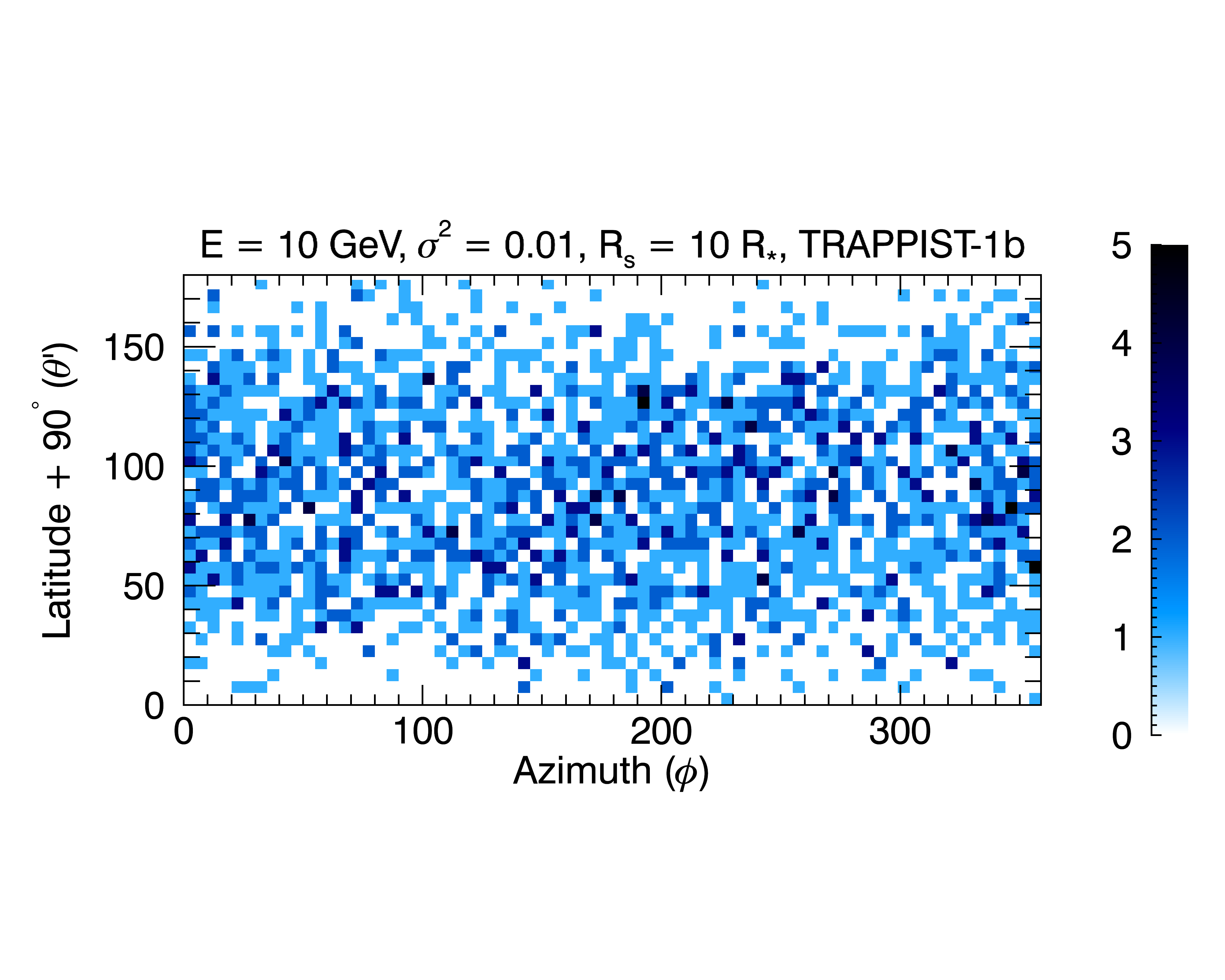}
	\includegraphics[width=9.1cm]{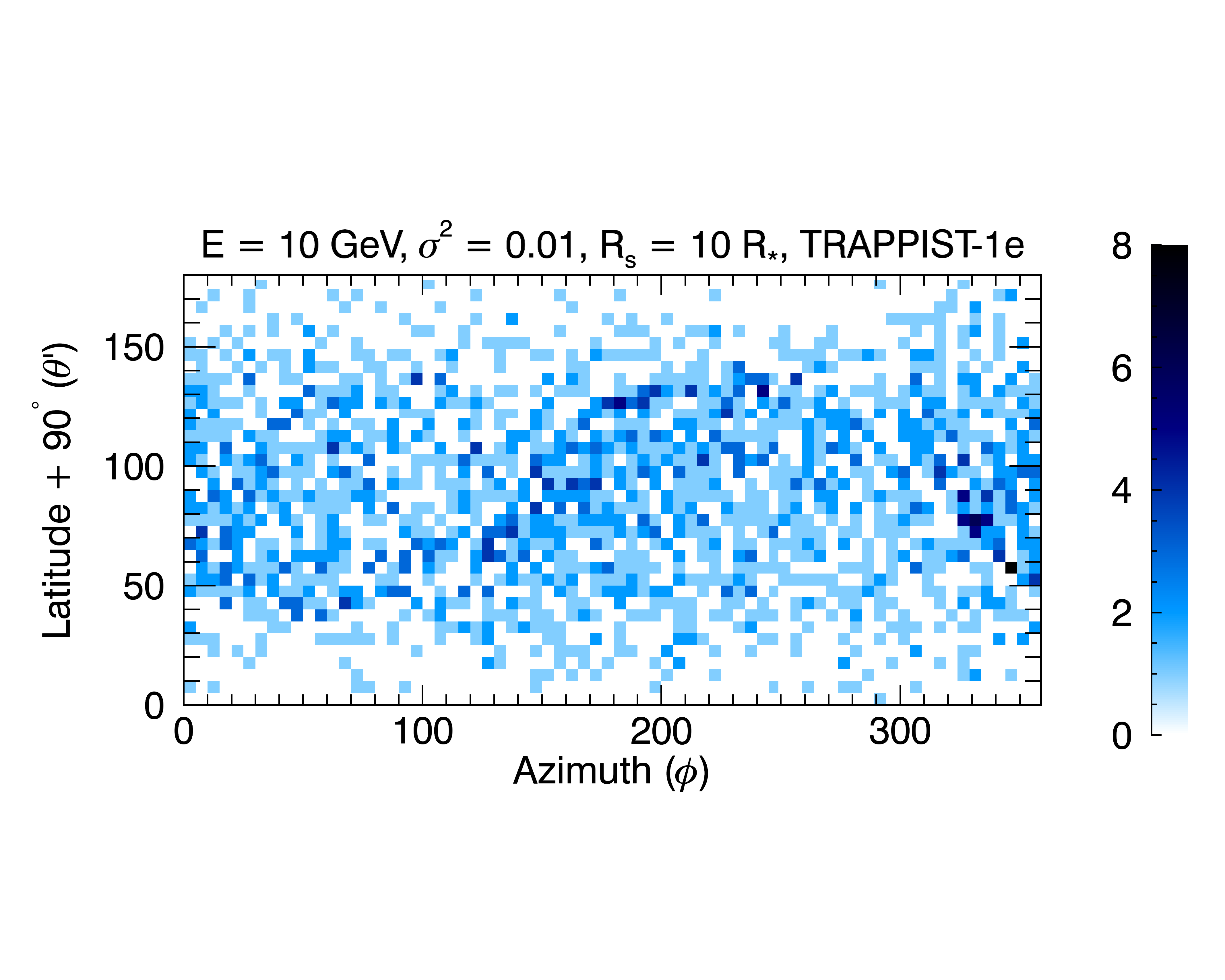}\\
	\includegraphics[width=9.1cm]{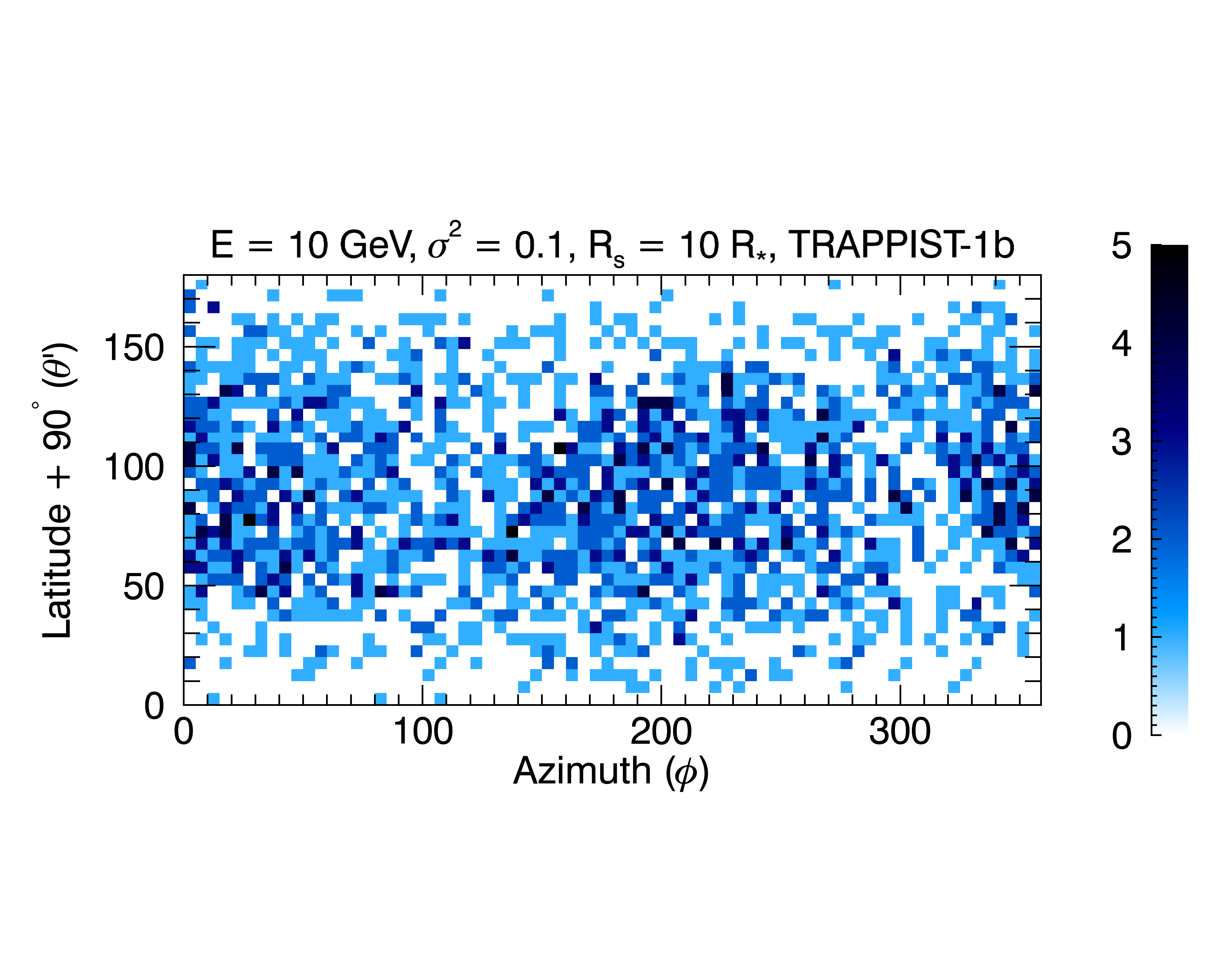}
	\includegraphics[width=9.1cm]{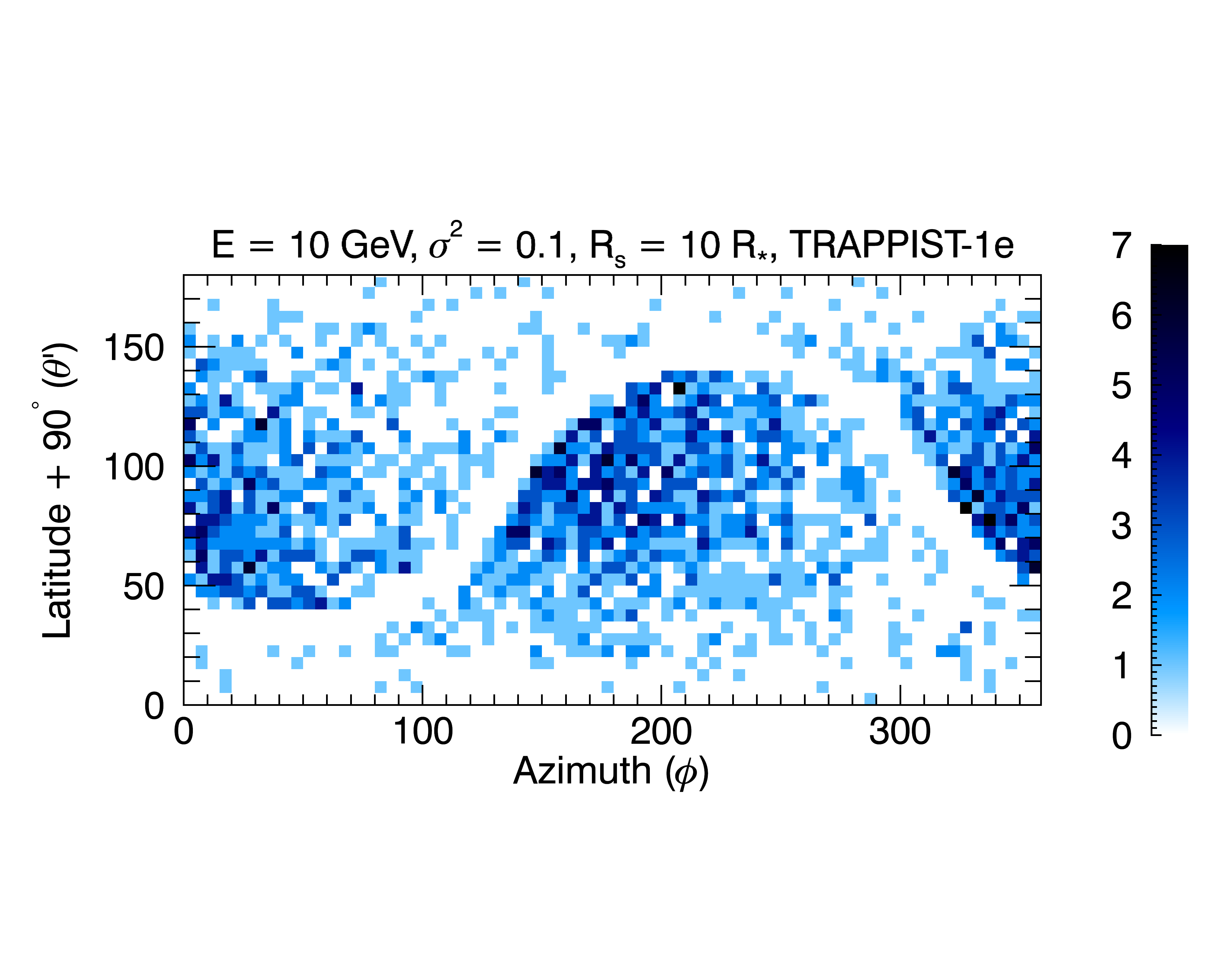}\\
	\includegraphics[width=9.1cm]{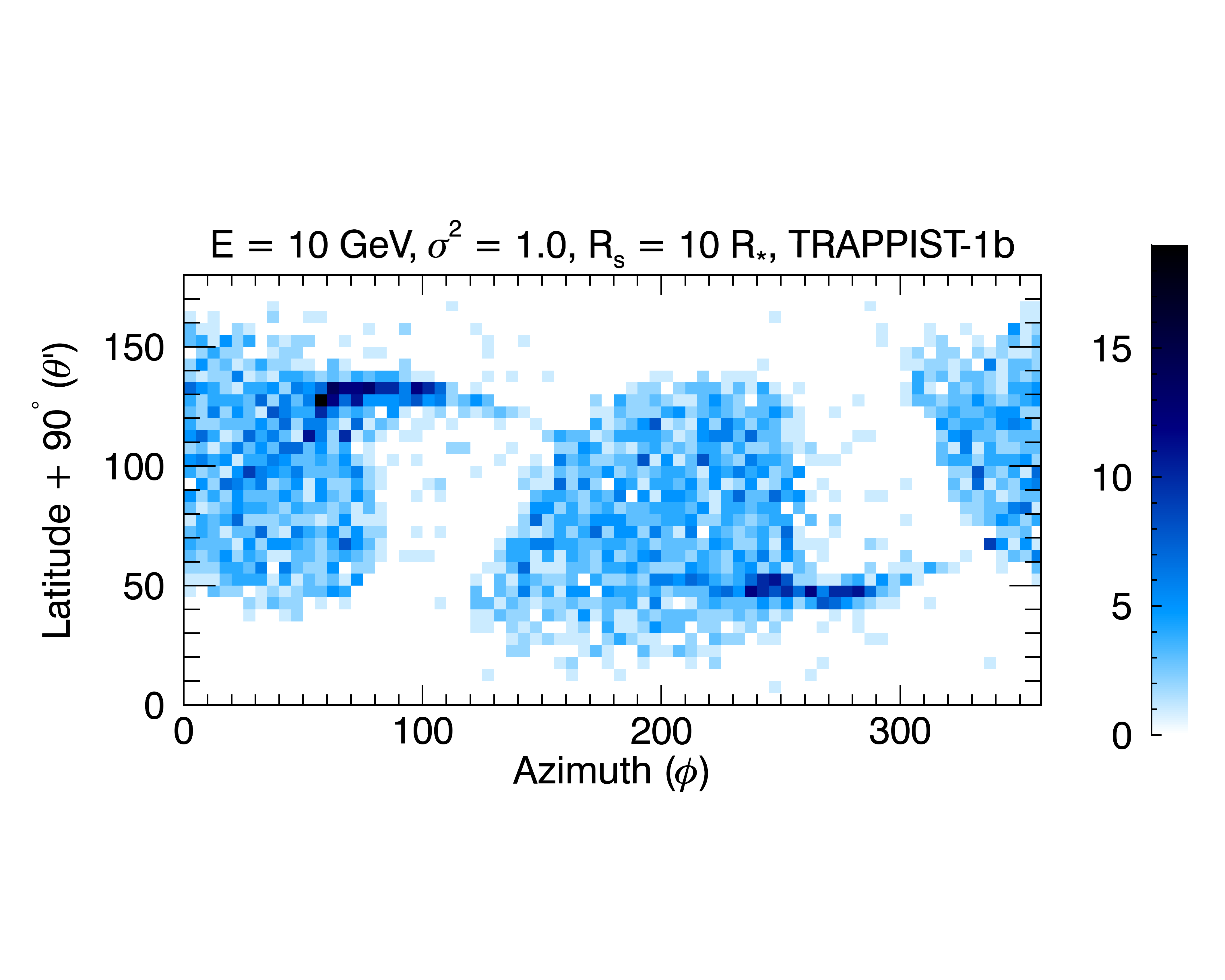}
	\includegraphics[width=9.1cm]{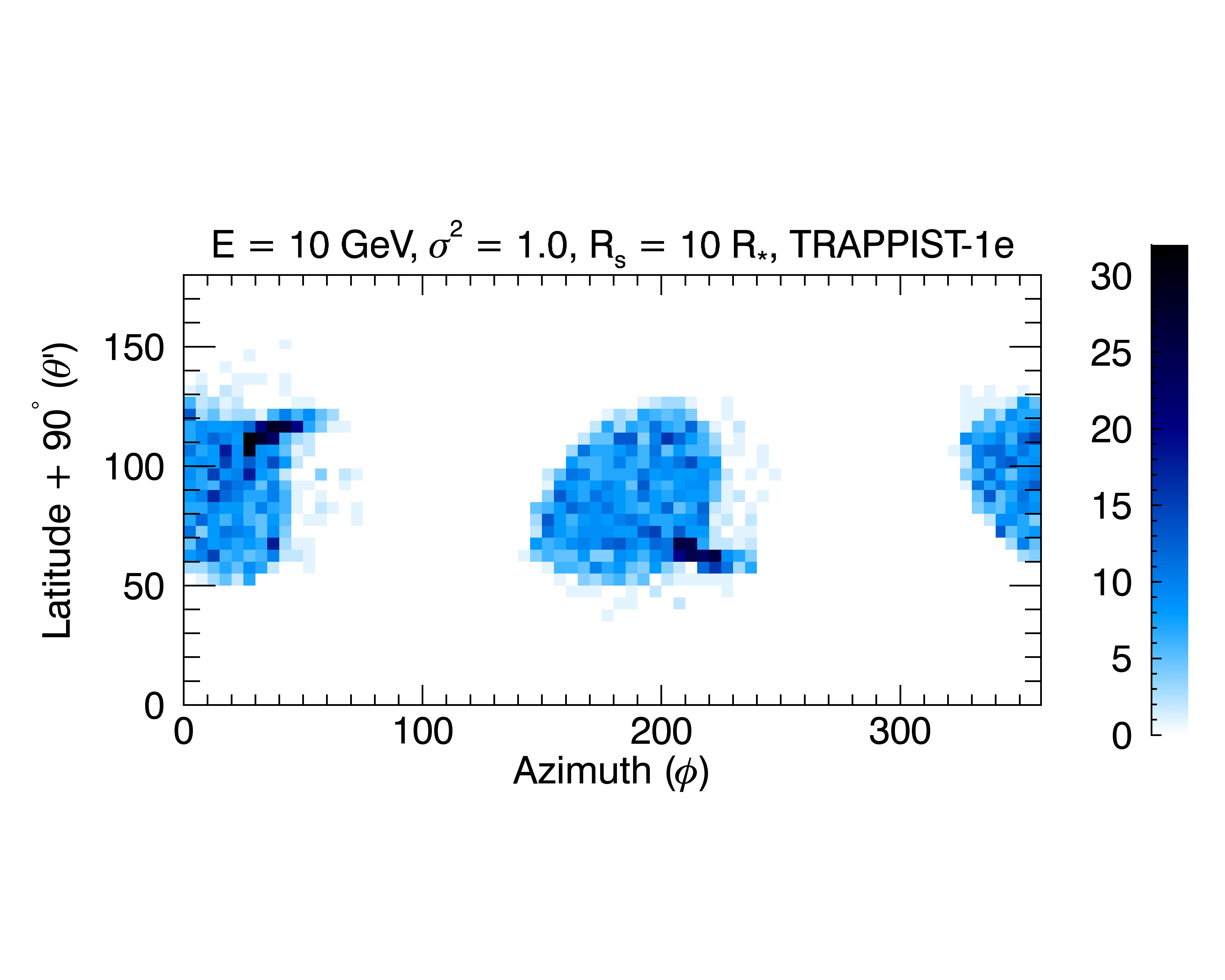}\\
\caption{Same as Fig. \ref{2D_E_1d3} for 10~GeV kinetic energy protons. \label{2D_E_1d4}}
\end{figure*}

\begin{figure*}
	\includegraphics[width=9.1cm]{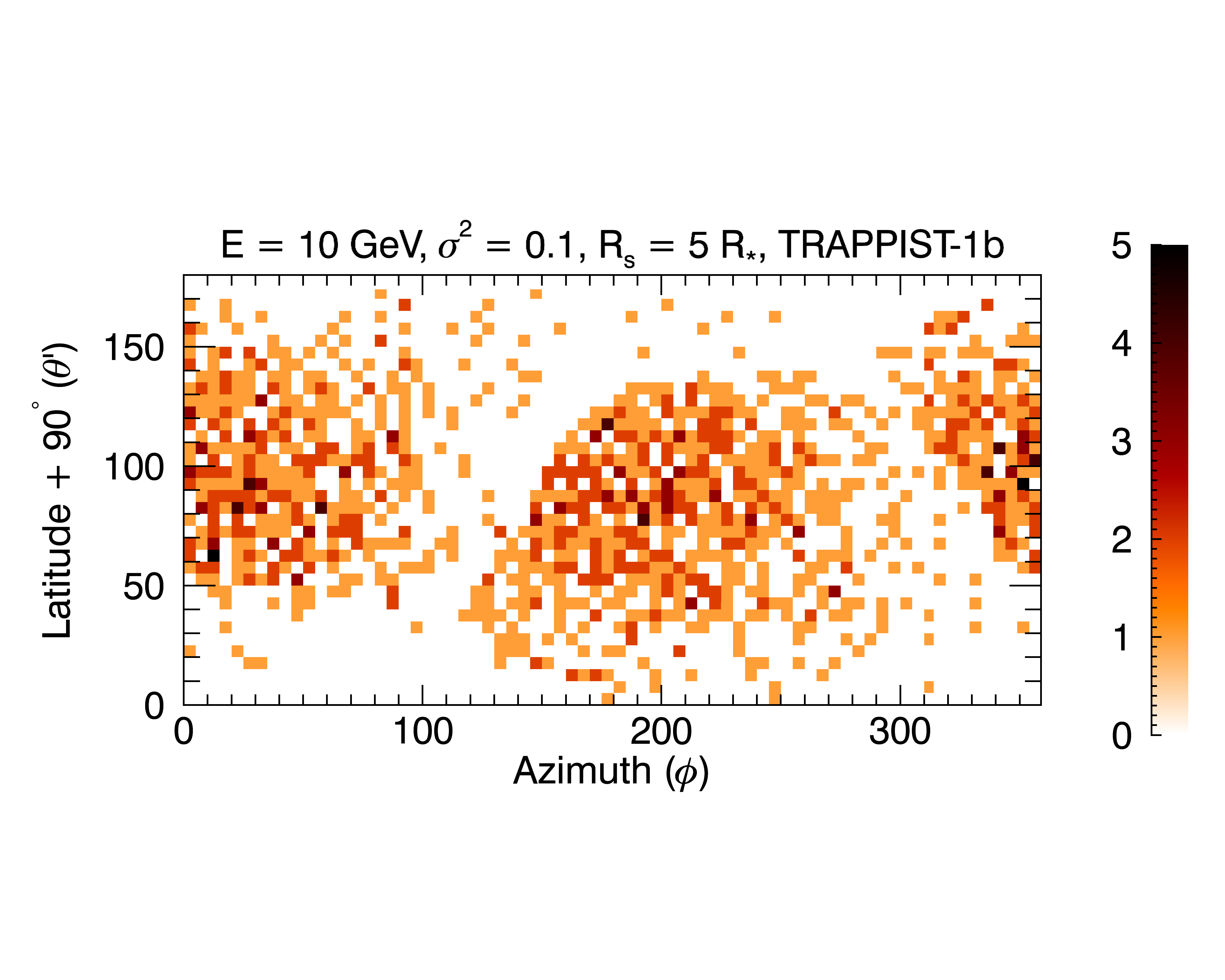}
	\includegraphics[width=9.1cm]{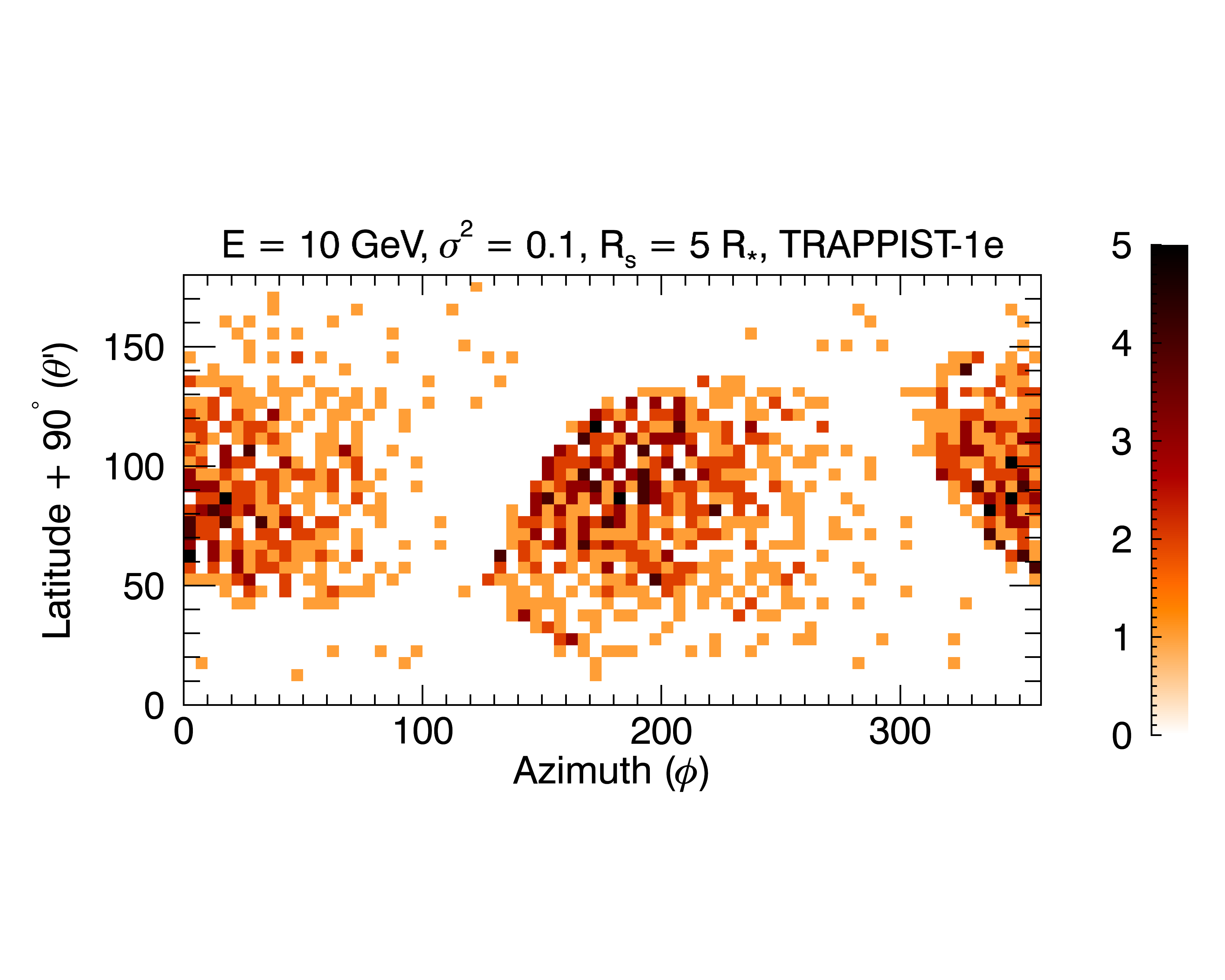}\\
\caption{Coordinates of the hitting points for 10~GeV kinetic energy protons, injected at $R_s = 5 R_{\star}$ with $L_c = 10^{-5}$~AU, at the spherical surface with radius $R_p$ equal to the semi-major axis of the planets TRAPPIST-1b (left column) and 1e (right column) and for $\sigma^2 = 0.1$. The same $x$ ($y$) axis and colorbar setting as in Fig. \ref{2D_E_1d3} are used.
\label{2D_E_1d4_5}}
\end{figure*}

\begin{figure*}
	\includegraphics[width=9.1cm]{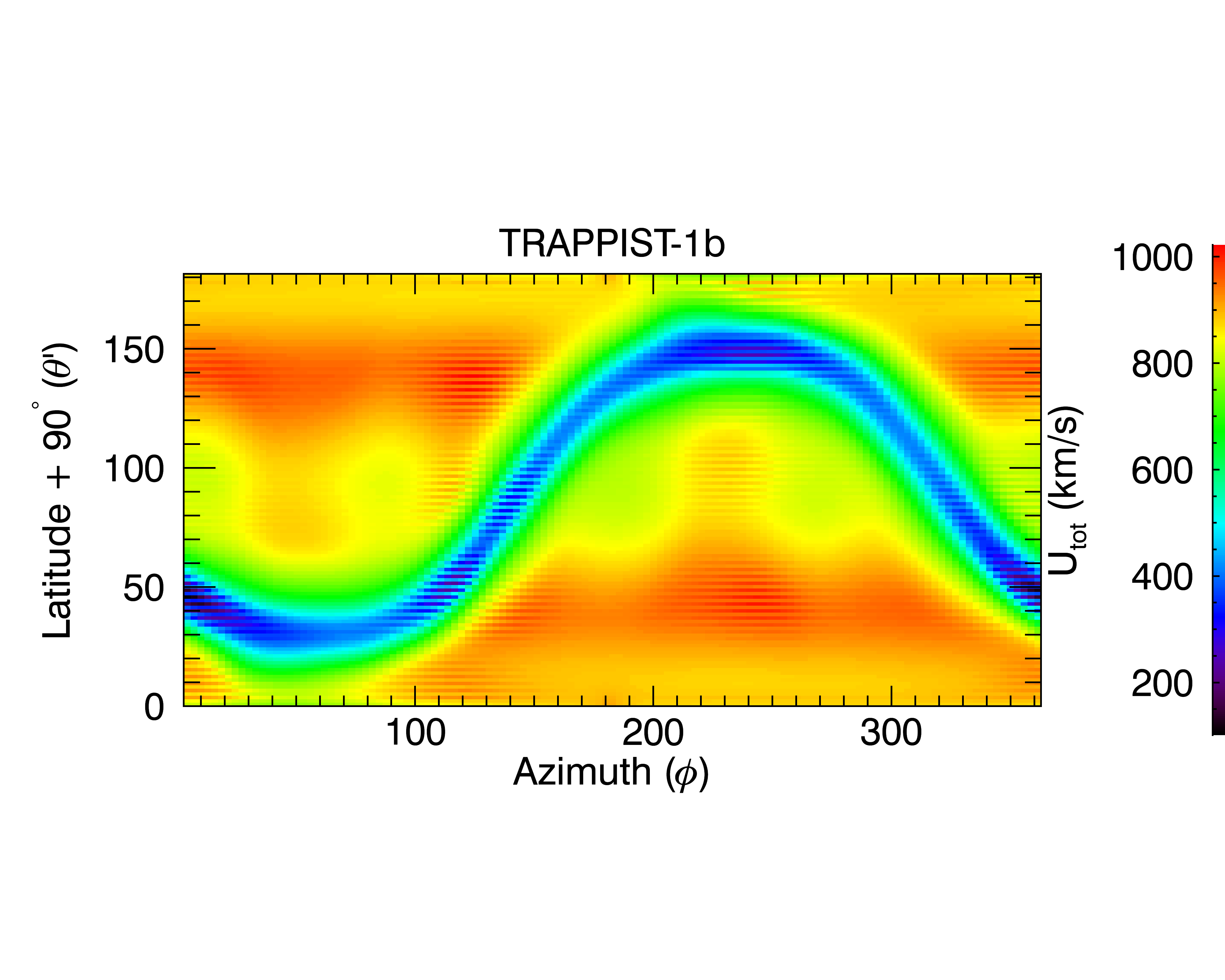}
	\includegraphics[width=9.1cm]{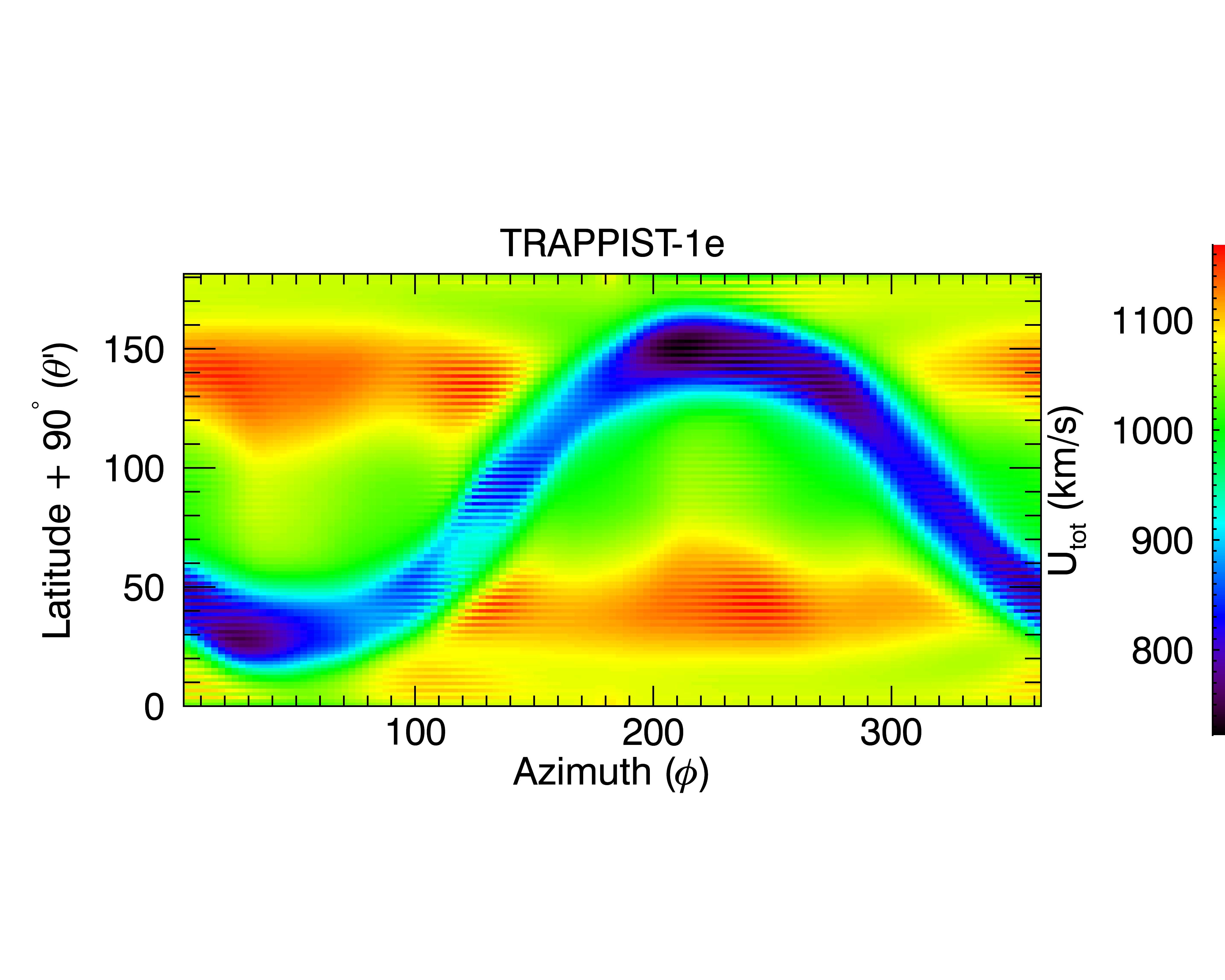}\\
	\includegraphics[width=9.1cm]{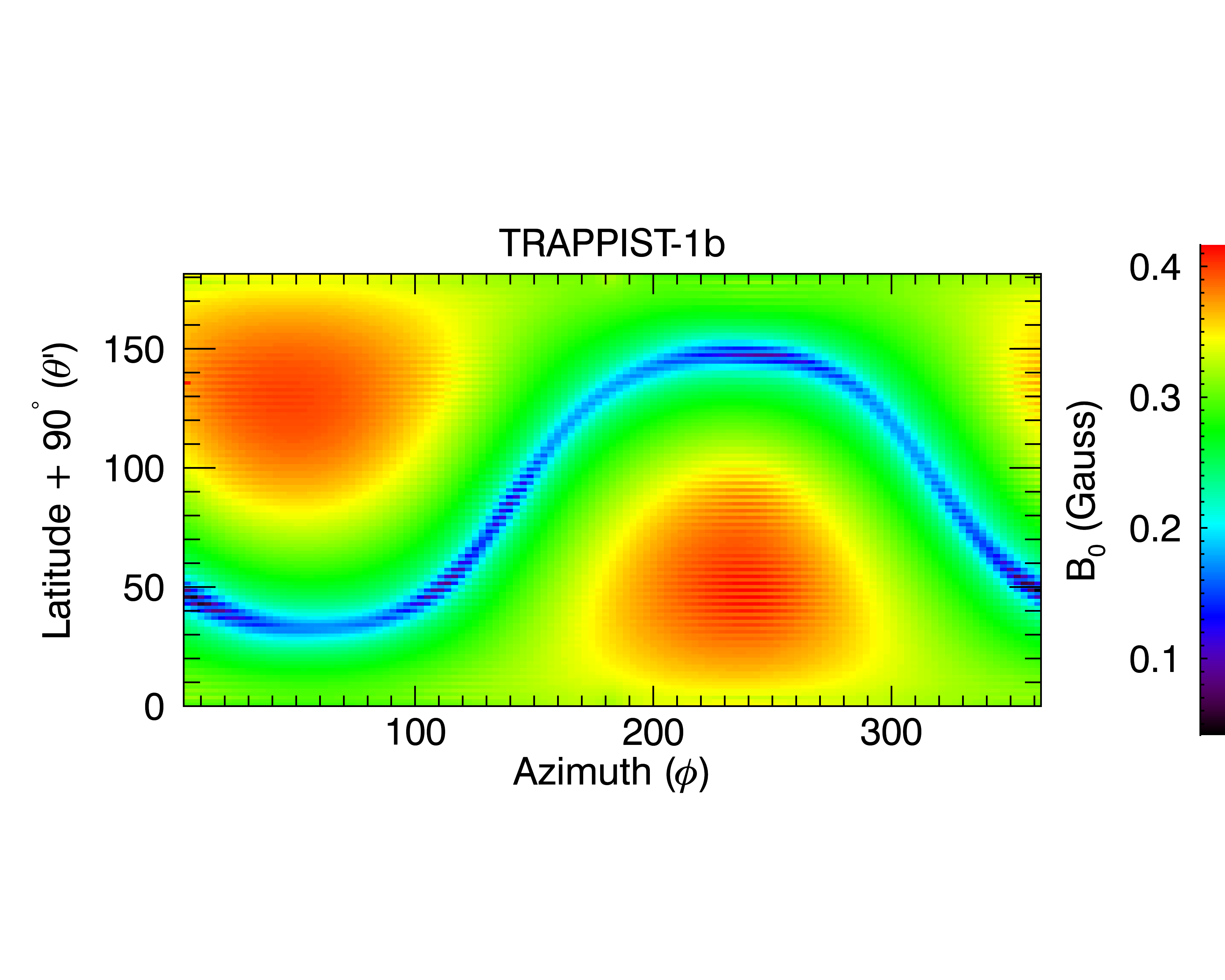}
	\includegraphics[width=9.1cm]{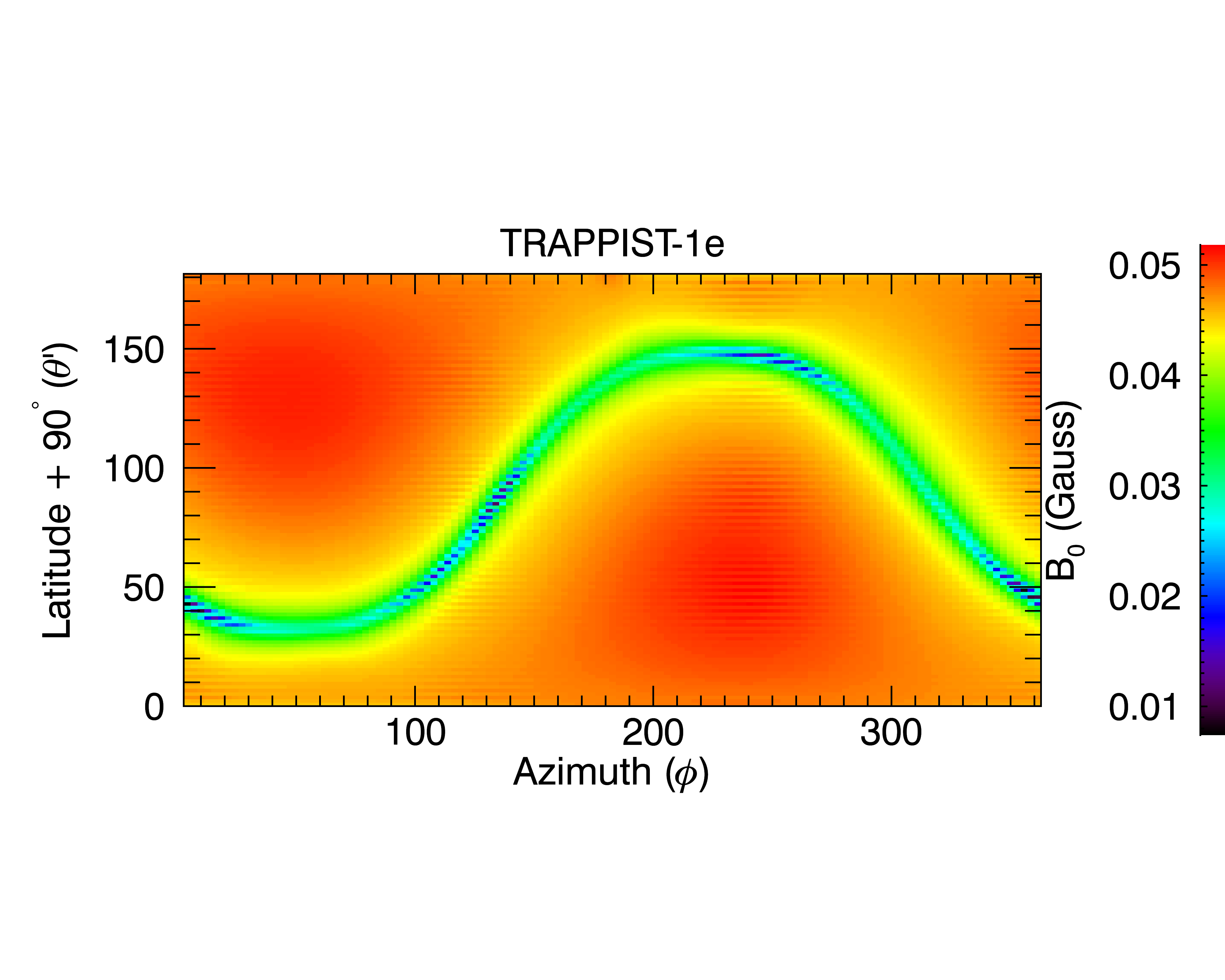}\\
\caption{Upper row: Magnitude of the total wind speed field $U$ on the $R_b$ (left) and $R_e$ (right) spherical surfaces. Lower row: Unperturbed magnetic strength $B_0$ on the $R_b$ (left) and $R_e$ (right) spherical surfaces. \label{2D_BU_1d4}}
\end{figure*}

The histogram on the $R_p$-sphere changes dramatically in the presence of stronger turbulence ($\sigma^2 = 0.1$, $1.0$, middle and lower row in Figs.~\ref{2D_E_1d3}, \ref{2D_E_1d4} and in Fig. \ref{2D_E_1d4_5}): EP hitting points on the $R_p$-sphere are confined to equatorial caps. We find a depleted region, in white, that is barely discernible at $R_p = R_b$ but conspicuous at $R_p = R_e$, and that azimuthally oscillates in the middle and bottom rows in Figs.~\ref{2D_E_1d3}, \ref{2D_E_1d4} and \ref{2D_E_1d4_5}.  This arises from the inclination of the magnetic axis to the rotation axis, and traces the azimuthal variation of the slow wind (see the spherical map of the wind speed, upper row in Fig.~\ref{2D_BU_1d4}). 

Inspection of the structure of the average magnetic field (see Fig.~\ref{f:windmag}) confirms that closed (open) field lines populate mainly the slow (fast) wind region. Moreover, a comparison of the middle row of Fig.~\ref{2D_E_1d4} with Fig. \ref{2D_E_1d4_5} shows that injection further out ($R_s = 10 R_*$ rather than $5 R_*$) reduces the chances of intercepting a closed field line due to the opening of field lines in the slow wind region as one proceeds outward. Consequently, the depleted white regions narrow down as the injection radius is increased from $R_s = 5$ to $10$. 

The broadening of the depleted regions as $\sigma^2$ increases, shown in the bottom rows of Figs.~\ref{2D_E_1d3} and \ref{2D_E_1d4}, can be explained as follows.
A greater amplitude of magnetic fluctuation, i.e.,  greater $\sigma^2$, leads to a reduced $\lambda_\parallel$ (see Eq. \ref{lambda}) and to an enhanced perpendicular diffusion: EPs more frequently decorrelate via cross-field transport. Near the boundary between open and closed field lines, a fraction of particles diffusing from open onto closed field will collapse back to the star, depleting the region corresponding to the current-sheet. There is then a net migration from open to closed field due to this loss of particles at the stellar surface.

The diffusive motion in the opposite direction, i.e., from a closed field lines near the boundary to an open line, and subsequent escape is less likely due to smaller $B_0$ of the closed line regions (see Fig.~\ref{2D_BU_1d4}, lower row), i.e. larger $\lambda_\parallel$, that might lead EPs rapidly to the stellar surface. Indeed, EPs can travel a short distance before falling to the star as the path length of the closed field lines is only a few times $\lambda_\parallel$ (from Eq. \ref{lambda}, a 10~GeV proton at $R_s = 10 R_*$, with $r_g/L_c \sim 0.1$, for $\sigma^2 = 0.1$ has $\lambda_\parallel \simeq 3.3 \times 10^9$ cm $\simeq 0.5 R_*$ that increases outward as shown in Sec. \ref{sec:weak}). We note that for the case of weak turbulence ($\sigma^2 =0.01$, Figs.~\ref{2D_E_1d3} and \ref{2D_E_1d4}, upper row) the depleted regions seen at higher $\sigma^2$ are not visible on the $R_p$-sphere as on the spheres at $R_p = R_b, R_e$ the points intercepted by open field lines are approximately uniform and closed lines do not reach such distances.

As for the escaping EPs, once they are channelled into the fast wind region, the large $B_0$ (see Fig.~\ref{2D_BU_1d4}, lower row) keeps them confined and focussed toward the caps, where $B_0$ is larger and hence $r_g$ smaller. 

Particularly relevant to the influence of EPs on  planets in our simulated magnetic field configuration is the approximate symmetry of the caps (see Sect. \ref{sec:flux}) with respect to the equatorial plane ($\theta' = 90^\circ$); such a pattern results within the fast wind region from the approximately symmetric and greater $B_0$ (lower row in Fig.~\ref{2D_BU_1d4}) that reduces $r_g$ thus favouring the confinement and focussing EPs within the caps. 

In the case of a Sun-like $B_0$-field, i.e., approximate alignment of ${\bf B}_0$ with the rotation axis, with $\sigma^2 \simeq 1$ (within the solar system typically $\sigma^2 < 0.1$), EPs would be directed preferentially into the polar regions, leaving planets relatively unaffected.
The solar wind latitudinal dependence of EPs in large events is, however, poorly constrained due to the limited number of events with high-latitude {\it in-situ} measurements (see Sect. \ref{sec:Discussion}).  

\begin{figure}
	\includegraphics[width=8.5cm]{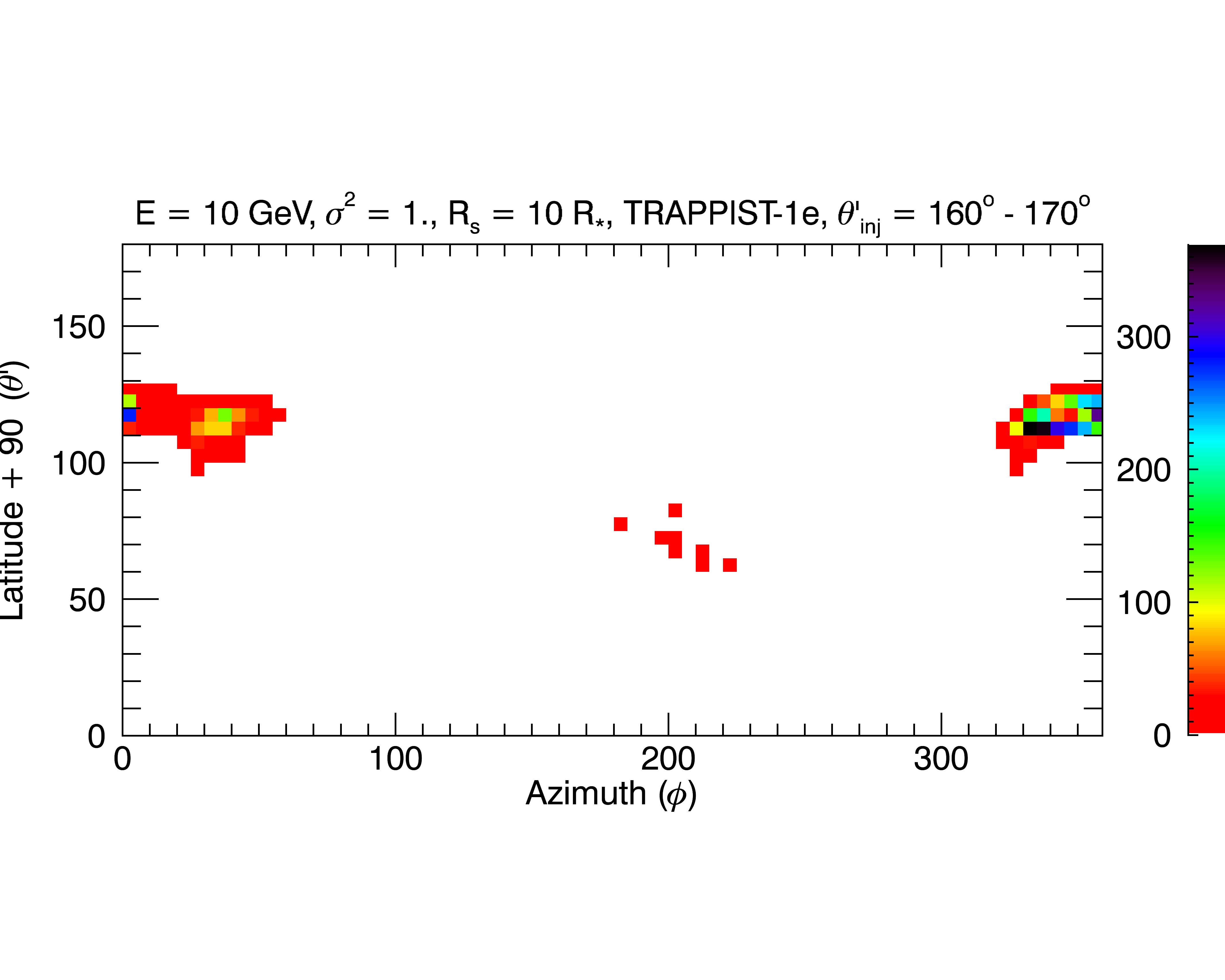}\\
\caption{Coordinates of the hitting points for 10~GeV kinetic energy protons, injected at $R_s = 10 R_{\star}$ on the latitudinal ring within the range $\theta' = 160^\circ - 170^\circ$ at the sphere with $R_p = R_e$ and for $\sigma^2 = 1.$. The $x$ ($y$) axis indicates the azimuthal (polar) coordinates on that sphere. The colorbar measures the number of EPs relative to the maximum. \label{2D_E_1d4_lat}}
\end{figure}

Surprisingly we find that EPs are focussed toward the equatorial plane even when injected at high latitude, i.e., close to the pole. Such an effect is shown in Fig. \ref{2D_E_1d4_lat} where EPs are injected, with isotropic velocity distribution, in the latitudinal ring in the upper hemisphere close to the geographic north pole with $\theta ' = 160 -170^\circ$. In this case, EPs are focused on the $R_p$-sphere within $40^\circ$ from the equatorial plane mostly in the upper hemisphere, except for a few points in the lower hemisphere ($180^\circ < \phi < 230^\circ$) due to an additional  diffusion in the azimuthal direction. 

\begin{figure}
    \hspace{-0.5cm}
	\includegraphics[width=4.8cm]{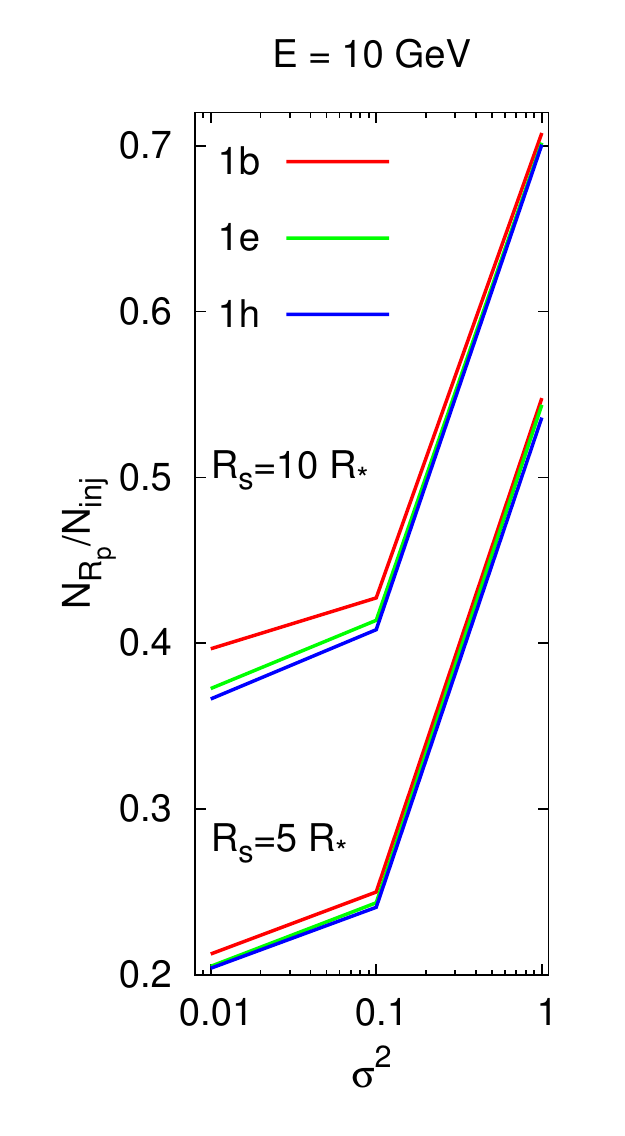} \hspace{-1.cm}
\includegraphics[width=4.8cm]{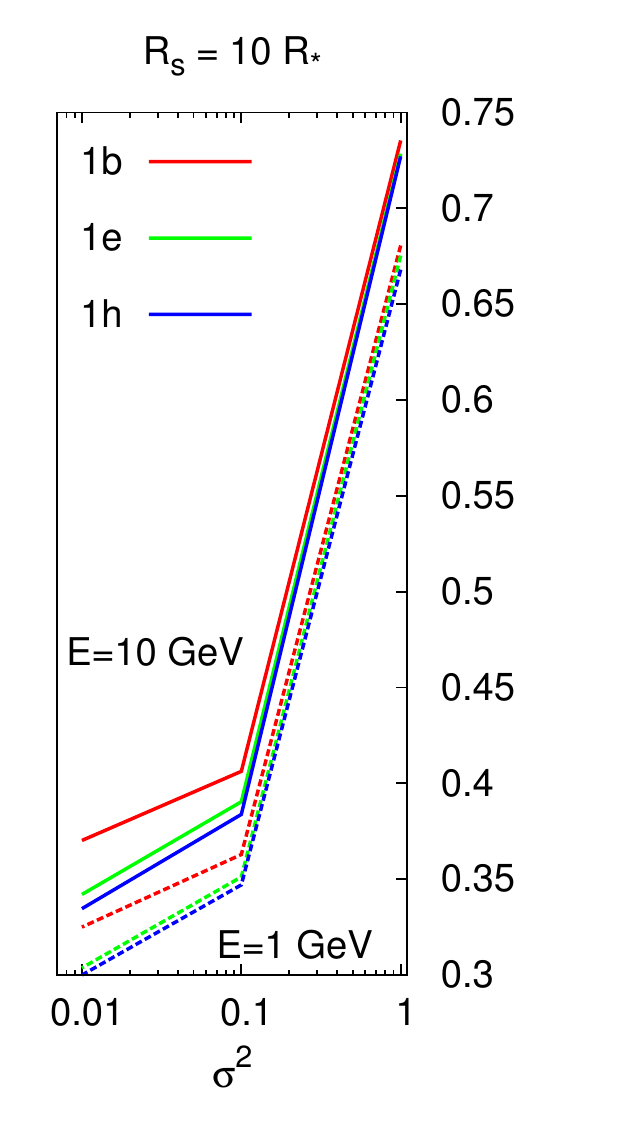}
\caption{Left: Fraction of EPs hitting the $R_p$-sphere for planets 1b (red), 1e (green), 1h (blue) relative to the total injected EPs as a function of $\sigma^2$, for $10$ GeV protons injected at $R_s =5$ and $10 R_*$. Right: Fraction of EPs hitting the $R_p$-sphere (same color legenda as left panel) relative to the total injected EPs as a function of $\sigma^2$, for $10$ GeV (solid) and $1$ GeV (dashed) protons injected, with equal $N_{inj}$, at $R_s = 10 R_*$  \label{Fraction_E_1d4}}
\end{figure}

We note that, despite the reduced filling factor of the EP caps for greater values of $\sigma^2$ shown in Figs.~\ref{2D_E_1d3} and \ref{2D_E_1d4}, that would seem to suggest a smaller $N_{R_p}$, the ratio $N_{R_p}/N_{\rm inj}$ actually {\em increases} for greater $\sigma^2$ (see Fig. \ref{Fraction_E_1d4}). This effect results again from (1) a more efficient perpendicular diffusion at the boundary between open and closed field lines and from (2) the increase of $\lambda_\parallel$ with distance from the star ($\lambda_\parallel \propto B_0 (r)^{-1/3}$). For {\it most} EPs injected on open field lines near the boundary, the former enhances the frequency of decorrelation from a given field line, as discussed above, and the latter favours EPs moving outward with an increasing $\lambda_\parallel$ rather than back to the star. Such combined effects ultimately prevent most particles from collapsing to the star and allow them to propagate outward toward the equatorial caps.

At larger EP energy, the escape of EPs injected at the open/closed field line boundary is favoured, as suggested by Fig.~\ref{Fraction_E_1d4}, right panel: $10$ GeV protons arrive more copiously on the $R_p$-spheres than $1$ GeV ones. This is a result of a larger perpendicular transport coefficient at larger energy, regardless of the particular model.

Finally, the features in the bottom rows of Fig.~\ref{2D_E_1d4} protruding out of the caps toward greater $\phi$, and also present to a lesser extent in Fig.~\ref{2D_E_1d3}, map the stripe at constant latitude of maximal wind flow visible in red in Fig.~\ref{2D_BU_1d4}, lower panels. On the other hand, the EP caps are shifted to smaller $\phi$ as a result of the stellar rotation. 

\section{Energetic Particle Flux within the TRAPPIST-1 System}\label{sec:flux}

The total output of EPs from M dwarf stars cannot be measured directly at present. A possible approach to estimate the EP abundance relies on the solar correlations between the observed properties of coronal flares and {\it in-situ} spacecraft measurements of EP fluxes at $1$ AU. GOES observations of $800$ SXR solar flares ($1.5-12.4$ keV) at the Sun and measurements of the associated $> 10$ MeV energetic protons events have shown an approximately linear correlation of the far-UV emission line flux to the proton flux \citep{Belov.etal:07}. 

\cite{Youngblood.etal:17} found two correlations: (1) between SXR peak flux and the flux of $> 10$ MeV protons from GOES data only; (2) between SDO/EVE He II 304 $\AA$ emission line fluence during the entire durations of flares and $> 10$ MeV GOES protons fluence. By using a sample of stellar flares observed by the Hubble Space Telescope (HST) and {\it Chandra}/ACIS, \cite{Youngblood.etal:17} finally inferred the proton enhancement for other stars. The He II 304 $\AA$ ($41$ eV) flare fluence was related to the HST far-UV ($7.3 - 13.6$ eV) fluence with the M dwarf synthetic spectrum created in \cite{Fontenla.etal:16}. The solar flaring rates for M- and X-class (corresponding to a SXR peak flare flux of $10^{-5}$ and $10^{-4}$ W/m$^{2}$ at $1$ AU in the $[1-8] \, \textup{\AA}$ band in the GOES classification, respectively) are estimated to be $0.02$ hr$^{-1}$ and $2.3 \times 10^{-5}$ hr$^{-1}$, respectively, based on flare observations in the period 1976-2000 \citep{Veronig.etal:02}. Therefore, the estimated rates for M- and X-class flares on the M4 dwarf GJ 876 are $\sim 0.4$ hr$^{-1}$ \citep{Youngblood.etal:17}, $20$ and $1.7 \times 10^5$ times more frequent than the Sun for M- and X-class, respectively. The rescaling to the average HZ radius $r_{876}^{HZ} \sim 0.18$ AU \citep[via the empirical scaling in \cite{Kopparapu.etal:14}]{Youngblood.etal:17}, leads to an increase of the flux by a factor $30$ for the HZ of GJ 876 (a flaring rate $600$ and $5 \times 10^5$ times higher for M- and X-class, respectively); it should also be noted that, due to the closer HZ, M-class flares are scaled up to X10. Therefore, \citet{Youngblood.etal:17} estimate that large GJ 876 flares (SXR peak flux $ \geq 10^{-3}$ W m$^{-2}$) lead to a $> 10$ MeV proton flux ($F_{876}^{max}$) on HZ planets up to $10^3$ protons cm$^{-2}$ s$^{-1}$ sr$^{-1}$, and enhanced up to $\sim 4$ orders of magnitude higher than for the present-day Earth by both the higher flaring rate and closer distance. 

Since the \cite{Youngblood.etal:17} scaling applies to EPs of any energy $> 10$ MeV, it should be noted that here we implicitly assume a uniform EP energy spectrum, although different spectral shapes, e.g., power-law or log-parabola, normalized to $> 10$ MeV could be used. 

The TRAPPIST-1 HZ is dramatically closer to the host star ($R_e = 0.029$ AU) than the GJ 876 HZ, leading to a much higher EP flux. Rescaling the flux from $r_{876}^{HZ} = 0.18$ AU to the injection radius in our simulations, $R_s = 10 R_* = 0.0056$ AU, we find an EP flux enhancement 
\begin{equation}
F_{\rm inj} (R_s) =  \left(\frac{r_{876}^{HZ} }{ R_s}\right)^2 F_{876}^{max} \simeq 10^3 \times F_{876}^{max} \simeq 10^6 \, \frac{{\rm protons}}{{\rm cm}^2 \, {\rm s} \,  {\rm sterad}} \, .
\label{max_flux}
\end{equation}
The relation above holds for very intense flares.

By using the maximal EP flux in Eq.~\ref{max_flux}, we can determine the flux $F (R_p)$ of EPs impinging on the planet 1e along its 6 day orbital motion around the star. The EP flux impinging on a ring of the $R_p$-sphere with semi-aperture $\Delta \theta' = 5^\circ$ centered on the equatorial plane is given by
\begin{equation}
F (R_p) = \frac{N'_{R_p}}{N_{\rm inj}}\frac{F_{\rm inj} (R_s)}{A} 
\label{D}
\end{equation}
where $N'_{R_p}$ is the number of EPs hitting the ring and we have used $ A = \int_{85^\circ}^{95^\circ} {\rm sin}\, \theta' d\theta' = 0.17$.

\begin{figure}
    \hspace{-0.5cm}
	\includegraphics[width=9.1cm]{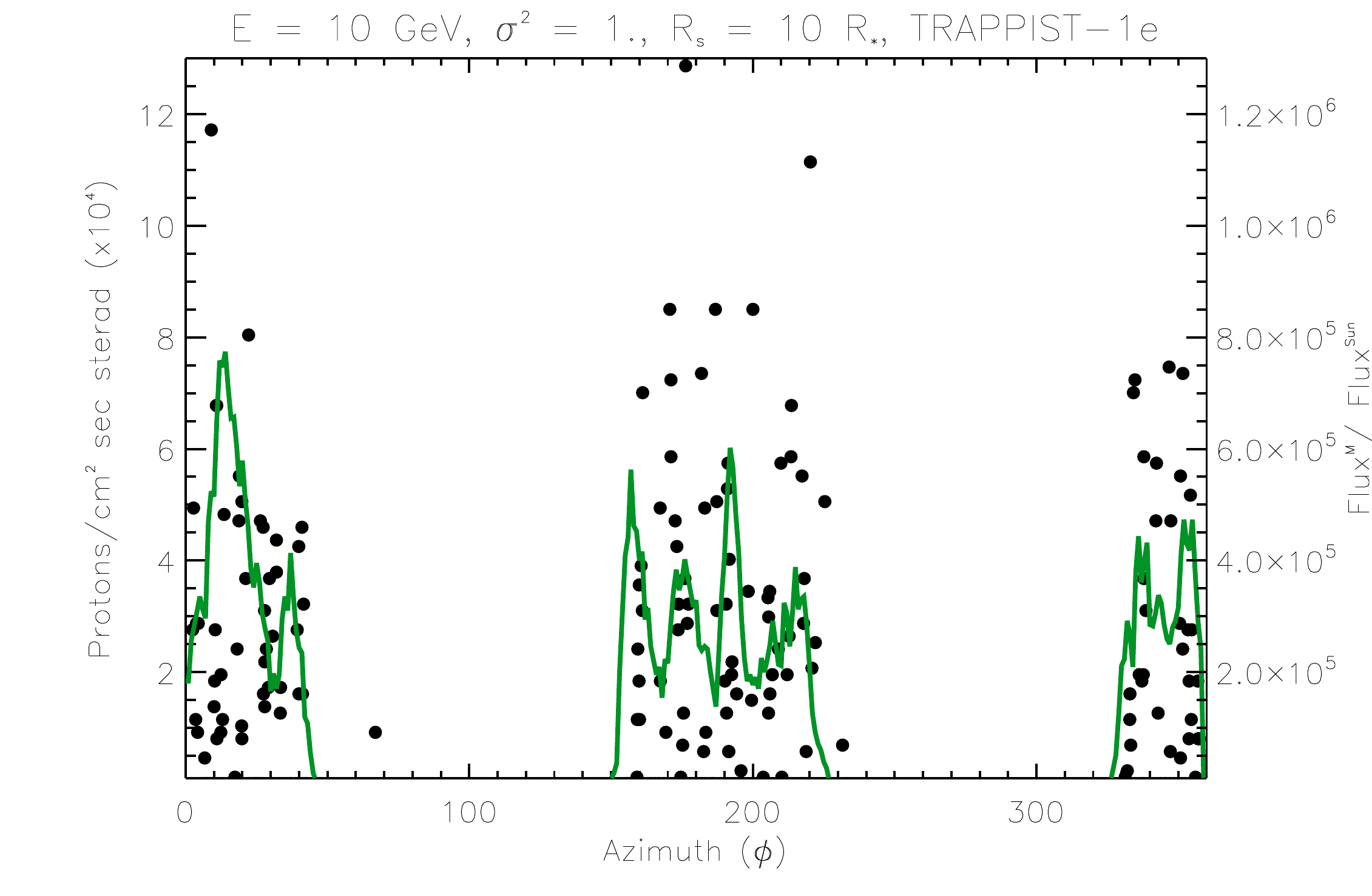}
\caption{Flux of $10$ GeV protons impinging onto a latitudinal ring of $5^\circ$ degrees semi-aperture centered on the equatorial plane for $R_p = R_e$, corresponding to the bottom row, right panel in Fig. \ref{2D_E_1d4}. Each point represents the total EP flux with an azimuthal binning of $1^\circ$. The green overlayed curve is the smoothed average using a $5^\circ$ boxcar smoothing width. The right hand side axis uses a very approximate renormalization to the solar EPs flux based on flaring rate estimate (see Sect. \ref{sec:flux}).  \label{1D_E_1d4_timevar}}
\end{figure}

The flux of $10$ GeV EPs with $\sigma^2 =1$, $R_s = 10 R_*$ along the orbit of planet $1$e is shown in Fig. \ref{1D_E_1d4_timevar}. 
The maximal flux, $\sim 1.2 \times 10^5 \, \frac{{\rm protons}}{{\rm cm}^2 \, {\rm s} \,  {\rm sterad}}$, exceeds by roughly $6$ orders of magnitude the EP abundance at the present-day Earth. However, such an estimate is subject to several caveats, which we discuss in the following section.

\section{Discussion}\label{sec:Discussion}

The results described in Sect.~\ref{sec: results} show that the magnetic fluctuations not only affect the small-scale particle motion but change drastically the behaviour of EPs over the entire inner astrosphere. 

\subsection{The spatial distribution of propagating EPs}

The EP-depleted angular regions on the $R_p$-sphere track the slow wind populated by closed field lines that lead to EPs being trapped and lost due to their trajectories leading back to the stellar surface.  For relatively large values of $\sigma^2$, particles are lost due to enhanced perpendicular diffusion into the closed field region (see Fig.s \ref{2D_E_1d3} and \ref{2D_E_1d4}). The opening of the closed field lines further out results in the narrowing of the depleted regions for larger particle injection radii $R_s = 10 R_*$ as compared to $R_s = 5 R_*$ (see Fig. \ref{2D_E_1d4_5}).

The stronger unperturbed magnetic field in the fast wind region on the equatorial plane (see Fig. \ref{2D_BU_1d4}, lower row) favours EP focussing. The EP caps are centered in the region of fast wind speed at $\sim 800-1,000$ ($\sim 950 - 1,100$) km$/$s at the planet 1b (1e).

A key characteristic of the GJ~3622 proxy magnetogram we adopted for TRAPPIST-1 is its resemblance to a tilted dipole.
This gives rise to the focus of EPs at low latitudes, and into the planetary orbital plane. The location of the spherical caps of EPs  hitting the $R_p$-sphere has potentially important consequences for the energetic particle flux experienced by the planets in our TRAPPIST-1-like system (the TRAPPIST-1 planets themselves are all in coplanar orbits to within $30$ arcmin). It should be noted that the locations of the EP caps are subject to shift both along the orbital plane due to  differences in the stellar rotation and planetary orbital periods, and in latitude due to the evolution and probable cyclic behavior of the stellar surface magnetic field. Both times scales associated with these processes are much greater than the EP propagation time scale. We investigate the EP flux variation planets could experience below.

We also point out that the EP focussing onto planets seen in our simulations is not expected to occur in a stellar wind driven by a dipolar magnetic field closely aligned with the stellar rotation axis (such as the solar wind), 
where the wind is fast at high latitudes (see Fig. \ref{2D_BU_1d4}, lower row). Moreover, $\sigma^2$ might attain values greater than $0.1$ only in transients, such as CME-driven shocks, or corotating interaction regions. In-situ solar wind measurements following large solar flares ($> 10^{30}$~erg) do not strongly constrain the latitudinal dependence in EP intensity: for instance, in the Bastille day event \citep{Zhang.etal:03} Ulysses high heliolatitude EP  intensity, in the fast wind, was measured at $3.2$ AU distance from the Sun whereas lower latitude intensity, in the slow wind, was measured at a different distance ($1$ AU). 

The spatial distribution of EPs centered on the equatorial plane might raise the question of a possible relation with the spatial distribution of CMEs in active M dwarfs found in numerical simulations by \cite{Kay.etal:16}: regardless of the latitude of injection, CMEs are deflected further out ($\sim 60 R_*$) along the near-equatorial current sheet, where the $B$-field is minimum and therefore CME expansion encounters the minimal magnetic confinement as the ratio between the CME ram pressure to the stellar magnetic pressure is highest. In our simulations, EPs are unleashed from the bulk motion of  CME-driven shocks at the initial time, so their motion is independent of the subsequent CME trajectory. We expect that in a stellar wind with a highly tilted magnetic-to-rotation axis, such as the one in Fig. \ref{f:windmag}, particles emitted at $R > 5 R_*$ by CMEs along the current sheet (blue-purple stripe in Fig. \ref{2D_BU_1d4}) will be transported toward the fast wind region for $\sigma^2 > 0.1$ (cfr. Fig. \ref{2D_E_1d4_lat}). However, for $\sigma^2 < 0.1$ we expect that the fewer escaping EPs will concentrate along the current-sheet stripe.

\subsection{On the absolute EP flux and trapping of EPs and CMEs}

Since EPs can be trapped by close field line regions, they can also be liberated from these regions when the closed field is perturbed or broken open. Such a disruption to the stellar magnetic $B_0$-structure can result from a CME-driven shock (not accounted for in our static solution MHD simulations), increasing the chances for EPs to fill the depleted regions on the $R_p$-sphere.

On the other hand, EPs accelerated and injected directly by coronal flares at $R_s < 2 R_*$, rather than by the travelling shock scenario considered in Figs.~\ref{2D_E_1d3},  \ref{2D_E_1d4}, \ref{2D_E_1d4_5},  are efficiently trapped by the very intense stellar magnetic field and by the closed field lines.  
Figure \ref{Fraction_E_1d4}, left panel, shows that doubling $R_s$ approximately doubles $N_{R_p}$.
The low $N_{R_p}/N_{\rm inj}$ ($3.0 - 3.7\%$) for $R_s = 1.5 R_*$
described in Sect. \ref{sec: results}, might be considered a lower limit if disturbances of the $B_0$ topology by flares or CMEs can enable a larger $N_{R_p}/N_{\rm inj}$.

These results indicate that a fairly simple dipole-like magnetic field structure on a magnetically active star prevents coronal flares from contributing significantly to the steady abundance of EPs further out. Thus, at face value in the undisrupted magnetic topology used here, CME-driven shocks might be expected to be the dominant supplier of EPs within the interplanetary medium of a very active star. 

In this context, the underlying assumption that CMEs can successfully escape the strong magnetic confinement of the stellar magnetic field to drive shock waves that accelerate EPs is uncertain and needs further investigation. \cite{Drake.etal:16} presented a preliminary simulation of what would have been a large CME on the Sun induced on the surface of the very active K dwarf AB~Dor, and found the event to be entirely contained by the strong overlying magnetic field.
Indication that a $75$ G dipolar field prevents the escape from the stellar corona of CMEs with kinetic energy $< 10^{32}$ erg has also been found by \cite{Alvarado.etal:18} based on a number of detailed numerical CME simulations.

There are thus two potentially powerful mechanisms that could strongly {\it limit} EP fluxes from active stars: EPs from flares are contained; and CMEs that might generate EPs at larger distances also fail to escape.

The morphology of $N_{R_p}/N_{\rm inj}$ in Figs.~\ref{2D_E_1d3}, \ref{2D_E_1d4} are, to a good approximation, independent of the EPs energy. In addition, the \cite{Youngblood.etal:17} correlation  is determined for $> 10$ MeV protons, with an unspecified EP energy-dependence. 
Regardless of the specific shape, we expect EPs energy spectrum to decrease at larger energy; thus, the EP flux $\sim 10^5 \, \frac{{\rm protons}}{{\rm cm}^2 \, {\rm s} \,  {\rm sterad}}$ impinging on 1e (see Sect.~\ref{sec:flux} and Fig. \ref{1D_E_1d4_timevar}) will be lower at $\gg 10$ MeV. We will investigate this effect in a forthcoming work.

We emphasize that our estimated number of injected EPs (Sect~\ref{sec:flux}) is based on strong flares in SXR observed from GJ 876 and classified as large, i.e., time-integrated SXR flux larger than $10^{29} - 10^{30}$ erg, due to the small distance to the star. The extrapolation of the correlation between SXR and EP fluence to such large events is uncertain due to the scatter of the observations and to the fact that no solar events beyond a certain energy have been observed \citep[$>$ X10, ][]{Hudson:07,Drake.etal:16}. However, {\it Kepler}-2 constraints \citep{Vida.etal:17} on TRAPPIST-1 white light flares lead to an estimated total  flare energy (in the optical) between $10^{31}$ and $10^{33}$ erg, similar to other very active M dwarfs \citep{Hawley.etal:14} and beyond the total estimated energy of the Carrington event \citep[$10^{32}$~erg, ][]{Carrington:1859} that is among the most energetic geomagnetic storms ever recorded on Earth. Thus, we argue that the dramatic EP enhancement in the HZ of M dwarfs like TRAPPIST-1 or GJ~876, as compared to present-day Earth, might be not uncommon.  Such EP fluxes could have a significant impact on exoplanet atmospheric ionization.

We do not consider the spatial distribution of the EP hitting points on the planetary surface or through the planetary atmosphere, since they depend strongly on the propagation through the planetary magnetosphere and atmosphere: the magnetospheric properties of the TRAPPIST-1 HZ planets---or any other exoplanets---are at present unknown. The effect of EPs on the atmospheric evolution also depends on the atmospheric mass and chemical composition, which are also unknown for TRAPPIST-1. Lyman $\alpha$ detection of variability during transits (observed for planets 1b and 1c, but not 1e,  \citealt{Bourrier.etal:17}), could be useful for further atmospheric characterization, although more detailed constraints will likely have to await observations by next generation facilities.

By using preliminary 3D-MHD simulations here, we instead consider simply the geometrical flux impinging onto a latitudinal ring, centered on the equatorial plane. We have integrated fluxes over a $5^\circ$ semi-aperture, which is much broader than the dispersion of the planetary orbits, in order to obtain sufficient signal from our test particle results (see Fig. \ref{1D_E_1d4_timevar}).

\section{Conclusions}\label{sec:Conclusion}

We have carried out numerical test-particle simulations to calculate for the first time the propagation of stellar energetic particles through a realistic and turbulent magnetic field of an M dwarf star and its wind.  Our simulations have been tailored to a proxy for TRAPPIST-1, and we have investigated the flux of energetic particles throughout the habitable zone of the TRAPPIST-1 system to the outermost planet. Particle acceleration by flares close to the stellar surface and further out by CME-driven shocks is mimicked here by injecting particles at various distances from the star over the full sphere and with an isotropic velocity distribution. We highlight three important aspects of the results.

Particles injected close to the stellar surface, regardless of their energy, are trapped within the strong stellar magnetic field. In our simulations, only a 3--4\%\ of particles injected within half a stellar radius from the surface escape. The escaping fraction increases strongly with increasing injection radius: Particles accelerated further from the stellar surface have a much greater chance of escaping the closed stellar magnetic field. 

Particles are increasingly focussed and directed toward the equator and toward open field fast wind regions with increasing turbulence amplitude. This results from asymmetric perpendicular diffusion from stronger to weaker field regions. In our TRAPPIST-1 proxy, strong turbulence produces two concentrated polar streams $180^{\circ}$ apart of energetic particles in the fast wind region focussed on the planetary orbital plane, regardless of the angular location of the injection.  Based on the scaling relation between far-UV emission and energetic protons for solar flares by \citet{Youngblood.etal:17}, we estimate that the innermost putative habitable planet, TRAPPIST-1e, is bombarded by a proton flux up to 6 orders of magnitude larger than experienced by the present-day Earth. Such a bombardement of planets in this study is found to result largely from the  misalignment of the $B$-field/rotation axis assumed for the star-proxy. Since the exact  magnetic morphology and alignment of the magnetic field is currently unknown for TRAPPIST-1, and for M dwarfs in general, our results indicate that determination of these quantities for exoplanet hosts would be of considerable value for understanding their radiation environments.

The trapping of EPs produced close to the stellar surface suggests that particles directly accelerated in flares do not generally escape, and that the ambient energetic particle environment of planets is dominated by particles accelerated in CME shocks.  However, recent findings that CMEs can be strongly suppressed by strong stellar magnetic fields \citep{Drake.etal:16,Alvarado.etal:18} point to a consequent large uncertainty in our understanding of the EP fluxes that exoplanets around active stars sustain.


\acknowledgments

We thank the referee whose constructive comments helped to improve significantly the clarity of the presentation. We are very grateful to Drs~G.~Ballester, K.~France, M.~Lingam and A.~Loeb for comments. The work of FF was supported, in part, by Scholarly Studies Award 40488100HH00181 at the Harvard-Smithsonian Center for Astrophysics, by NASA under Grants NNX15AJ71G and 80NSSC18K1213, by NASA through Chandra Award Number $TM6-17001B$ issued by the Chandra X-ray Observatory Center, which is operated by the Smithsonian Astrophysical Observatory for and on behalf of NASA under contract NAS8-03060 and by NSF under grant 1850774. JJD was funded by NASA contract NAS8-03060 to the CXC and thanks the Director, Belinda Wilkes, and the CXC science team for continuing advice and support. JDAG was supported by Chandra grants AR4-15000X and GO5-16021X. Resources supporting this work were partially provided by the NASA High-End Computing (HEC) Program through the NASA Advanced Supercomputing (NAS) Division at Ames Research Center. 

\def \apss{{\it Astrophys.\ Sp.\ Sci.}}
\def \aj{{\it AJ}}
\def \apj{{\it ApJ}}
\def \apjl{{\it ApJL}}
\def \apjs{{\it ApJS}}
\def \araa{{\it Ann. Rev. A \& A}}
\def \prc{{\it Phys.\ Rev.\ C}}
\def \aap{{\it A\&A}}
\def \aaps{{\it A\&ASS}}
\def \mnras{{\it MNRAS}}
\def \physscr{{\it Phys.\ Scripta}}
\def \pasp{{\it Publ.\ Astron.\ Soc.\ Pac.}}
\def \gca{{\it Geochim. Cosmochim.\ Act.}}
\def \nat{{\it Nature}}
\def \solphys{{\it Sol.\ Phys.}}

\bibliographystyle{aasjournal}
\bibliography{ms}

\begin{thebibliography}{}
\expandafter\ifx\csname natexlab\endcsname\relax\def\natexlab#1{#1}\fi
\providecommand{\url}[1]{\href{#1}{#1}}

\bibitem[{{Airapetian} {et~al.}(2016){Airapetian}, {Glocer}, {Gronoff},
  {H{\'e}brard}, \& {Danchi}}]{Airapetian.etal:16}
{Airapetian}, V.~S., {Glocer}, A., {Gronoff}, G., {H{\'e}brard}, E., \&
  {Danchi}, W. 2016, Nature Geoscience, 9, 452

\bibitem[{{Alvarado-G{\'o}mez} {et~al.}(2018){Alvarado-G{\'o}mez}, {Drake},
  {Cohen}, {Moschou}, \& {Garraffo}}]{Alvarado.etal:18}
{Alvarado-G{\'o}mez}, J.~D., {Drake}, J.~J., {Cohen}, O., {Moschou}, S.~P., \&
  {Garraffo}, C. 2018, \apj, 862, 93

\bibitem[{{Alvarado-G{\'o}mez}
  {et~al.}(2016{\natexlab{a}}){Alvarado-G{\'o}mez}, {Hussain}, {Cohen},
  {Drake}, {Garraffo}, {Grunhut}, \& {Gombosi}}]{Alvarado.etal:16a}
{Alvarado-G{\'o}mez}, J.~D., {Hussain}, G.~A.~J., {Cohen}, O., {et~al.}
  2016{\natexlab{a}}, \aap, 588, A28

\bibitem[{{Alvarado-G{\'o}mez}
  {et~al.}(2016{\natexlab{b}}){Alvarado-G{\'o}mez}, {Hussain}, {Cohen},
  {Drake}, {Garraffo}, {Grunhut}, \& {Gombosi}}]{Alvarado.etal:16b}
---. 2016{\natexlab{b}}, \aap, 594, A95

\bibitem[{{Armstrong} {et~al.}(1995){Armstrong}, {Rickett}, \&
  {Spangler}}]{Armstrong.etal:95}
{Armstrong}, J.~W., {Rickett}, B.~J., \& {Spangler}, S.~R. 1995, \apj, 443, 209

\bibitem[{{Barnes} {et~al.}(2014){Barnes}, {Jenkins}, {Jones}, {Jeffers},
  {Rojo}, {Arriagada}, {Jord{\'a}n}, {Minniti}, {Tuomi}, {Pinfield}, \&
  {Anglada-Escud{\'e}}}]{Barnes.etal:14}
{Barnes}, J.~R., {Jenkins}, J.~S., {Jones}, H.~R.~A., {et~al.} 2014, \mnras,
  439, 3094

\bibitem[{{Belov} {et~al.}(2007){Belov}, {Kurt}, {Mavromichalaki}, \&
  {Gerontidou}}]{Belov.etal:07}
{Belov}, A., {Kurt}, V., {Mavromichalaki}, H., \& {Gerontidou}, M. 2007,
  \solphys, 246, 457

\bibitem[{{Bourrier} {et~al.}(2017){Bourrier}, {Ehrenreich}, {Wheatley},
  {Bolmont}, {Gillon}, {de Wit}, {Burgasser}, {Jehin}, {Queloz}, \&
  {Triaud}}]{Bourrier.etal:17}
{Bourrier}, V., {Ehrenreich}, D., {Wheatley}, P.~J., {et~al.} 2017, \aap, 599,
  L3

\bibitem[{{Burlaga} \& {Turner}(1976)}]{Burlaga.Turner:76}
{Burlaga}, L.~F., \& {Turner}, J.~M. 1976, \jgr, 81, 73

\bibitem[{{Carrington}(1859)}]{Carrington:1859}
{Carrington}, R.~C. 1859, \mnras, 20, 13

\bibitem[{{Cohen} {et~al.}(2010){Cohen}, {Drake}, {Kashyap}, {Korhonen},
  {Elstner}, \& {Gombosi}}]{Cohen.etal:10b}
{Cohen}, O., {Drake}, J.~J., {Kashyap}, V.~L., {et~al.} 2010, \apj, 719, 299

\bibitem[{{Dauphas} \& {Chaussidon}(2011)}]{Dauphas.Chaussidon:11}
{Dauphas}, N., \& {Chaussidon}, M. 2011, Annual Review of Earth and Planetary
  Sciences, 39, 351

\bibitem[{{Delrez} {et~al.}(2018){Delrez}, {Gillon}, {Triaud}, {Demory}, {de
  Wit}, {Ingalls}, {Agol}, {Bolmont}, {Burdanov}, {Burgasser}, {Carey},
  {Jehin}, {Leconte}, {Lederer}, {Queloz}, {Selsis}, \& {Van
  Grootel}}]{Delrez.etal:18}
{Delrez}, L., {Gillon}, M., {Triaud}, A.~H.~M.~J., {et~al.} 2018, \mnras, 475,
  3577

\bibitem[{{Donati} \& {Brown}(1997)}]{Donati.Brown:97}
{Donati}, J.-F., \& {Brown}, S.~F. 1997, \aap, 326, 1135

\bibitem[{{Dong} {et~al.}(2018){Dong}, {Jin}, {Lingam}, {Airapetian}, {Ma}, \&
  {van der Holst}}]{Dong.etal:18}
{Dong}, C., {Jin}, M., {Lingam}, M., {et~al.} 2018, Proceedings of the National
  Academy of Science, 115, 260

\bibitem[{{Drake} {et~al.}(2016){Drake}, {Cohen}, {Garraffo}, \&
  {Kashyap}}]{Drake.etal:16}
{Drake}, J.~J., {Cohen}, O., {Garraffo}, C., \& {Kashyap}, V. 2016, in IAU
  Symposium, Vol. 320, Solar and Stellar Flares and their Effects on Planets,
  ed. A.~G. {Kosovichev}, S.~L. {Hawley}, \& P.~{Heinzel}, 196--201

\bibitem[{{Ellison} {et~al.}(1981){Ellison}, {Jones}, \&
  {Eichler}}]{Ellison.etal:81}
{Ellison}, D.~C., {Jones}, F.~C., \& {Eichler}, D. 1981, Journal of Geophysics
  Zeitschrift Geophysik, 50, 110

\bibitem[{{Emslie} {et~al.}(2012){Emslie}, {Dennis}, {Shih}, {Chamberlin},
  {Mewaldt}, {Moore}, {Share}, {Vourlidas}, \& {Welsch}}]{Emslie.etal:12}
{Emslie}, A.~G., {Dennis}, B.~R., {Shih}, A.~Y., {et~al.} 2012, \apj, 759, 71

\bibitem[{{Feigelson} {et~al.}(2002){Feigelson}, {Garmire}, \&
  {Pravdo}}]{Feigelson.etal:02}
{Feigelson}, E.~D., {Garmire}, G.~P., \& {Pravdo}, S.~H. 2002, \apj, 572, 335

\bibitem[{{Finley} \& {Matt}(2018)}]{Finley.Matt:18}
{Finley}, A.~J., \& {Matt}, S.~P. 2018, \apj, 854, 78

\bibitem[{{Fontenla} {et~al.}(2016){Fontenla}, {Linsky}, {Witbrod}, {France},
  {Buccino}, {Mauas}, {Vieytes}, \& {Walkowicz}}]{Fontenla.etal:16}
{Fontenla}, J.~M., {Linsky}, J.~L., {Witbrod}, J., {et~al.} 2016, \apj, 830,
  154

\bibitem[{{France} {et~al.}(2016){France}, {Loyd}, {Youngblood}, {Brown},
  {Schneider}, {Hawley}, {Froning}, {Linsky}, {Roberge}, {Buccino},
  {Davenport}, {Fontenla}, {Kaltenegger}, {Kowalski}, {Mauas}, {Miguel},
  {Redfield}, {Rugheimer}, {Tian}, {Vieytes}, {Walkowicz}, \&
  {Weisenburger}}]{France.etal:16}
{France}, K., {Loyd}, R.~O.~P., {Youngblood}, A., {et~al.} 2016, \apj, 820, 89

\bibitem[{{Fraschetti}(2016{\natexlab{a}})}]{Fraschetti:16a}
{Fraschetti}, F. 2016{\natexlab{a}}, \pre, 93, 013206

\bibitem[{{Fraschetti}(2016{\natexlab{b}})}]{Fraschetti:16b}
---. 2016{\natexlab{b}}, ASTRA Proceedings, 2, 63

\bibitem[{{Fraschetti} {et~al.}(2018){Fraschetti}, {Drake}, {Cohen}, \&
  {Garraffo}}]{Fraschetti.Drake.etal:18}
{Fraschetti}, F., {Drake}, J.~J., {Cohen}, O., \& {Garraffo}, C. 2018, \apj,
  853, 112

\bibitem[{{Fraschetti} \& {Giacalone}(2012)}]{Fraschetti.Giacalone:12}
{Fraschetti}, F., \& {Giacalone}, J. 2012, \apj, 755, 114

\bibitem[{{Fraschetti} \& {Jokipii}(2011)}]{Fraschetti.Jokipii:11}
{Fraschetti}, F., \& {Jokipii}, J.~R. 2011, \apj, 734, 83

\bibitem[{{Garraffo} {et~al.}(2015){Garraffo}, {Drake}, \&
  {Cohen}}]{Garraffo.etal:15}
{Garraffo}, C., {Drake}, J.~J., \& {Cohen}, O. 2015, \apj, 813, 40

\bibitem[{{Garraffo} {et~al.}(2016){Garraffo}, {Drake}, \&
  {Cohen}}]{Garraffo.etal:16}
---. 2016, \apjl, 833, L4

\bibitem[{{Garraffo} {et~al.}(2017){Garraffo}, {Drake}, {Cohen},
  {Alvarado-G{\'o}mez}, \& {Moschou}}]{Garraffo.etal:17}
{Garraffo}, C., {Drake}, J.~J., {Cohen}, O., {Alvarado-G{\'o}mez}, J.~D., \&
  {Moschou}, S.~P. 2017, \apjl, 843, L33

\bibitem[{{Giacalone} \& {Jokipii}(1999)}]{Giacalone.Jokipii:99}
{Giacalone}, J., \& {Jokipii}, J.~R. 1999, \apj, 520, 204

\bibitem[{Gillon {et~al.}(2017)Gillon, Triaud, Demory, Jehin, Agol, Deck,
  Lederer, De~Wit, Burdanov, Ingalls, {et~al.}}]{Gillon.etal:17}
Gillon, M., Triaud, A.~H., Demory, B.-O., {et~al.} 2017, Nature, 542, 456

\bibitem[{{Goldreich} \& {Sridhar}(1995)}]{Goldreich.Sridhar:95}
{Goldreich}, P., \& {Sridhar}, S. 1995, \apj, 438, 763

\bibitem[{{Gregory} {et~al.}(2009){Gregory}, {Matt}, {Donati}, \&
  {Jardine}}]{Gregory.etal:09}
{Gregory}, S.~G., {Matt}, S.~P., {Donati}, J.-F., \& {Jardine}, M. 2009, in
  American Institute of Physics Conference Series, Vol. 1094, 15th Cambridge
  Workshop on Cool Stars, Stellar Systems, and the Sun, ed. E.~{Stempels},
  71--76

\bibitem[{{Hawley} {et~al.}(2014){Hawley}, {Davenport}, {Kowalski},
  {Wisniewski}, {Hebb}, {Deitrick}, \& {Hilton}}]{Hawley.etal:14}
{Hawley}, S.~L., {Davenport}, J.~R.~A., {Kowalski}, A.~F., {et~al.} 2014, \apj,
  797, 121

\bibitem[{{Horbury} {et~al.}(2008){Horbury}, {Forman}, \&
  {Oughton}}]{Horbury.etal:08}
{Horbury}, T.~S., {Forman}, M., \& {Oughton}, S. 2008, Physical Review Letters,
  101, 175005

\bibitem[{{Horbury} \& {Tsurutani}(2001)}]{Horbury.Tsurutani:01}
{Horbury}, T.~S., \& {Tsurutani}, B. 2001, {Ulysses measurements of waves,
  turbulence and discontinuities}, ed. A.~{Balogh}, R.~G. {Marsden}, \& E.~J.
  {Smith}, 167--227

\bibitem[{{Hudson}(2007)}]{Hudson:07}
{Hudson}, H.~S. 2007, \apjl, 663, L45

\bibitem[{{Jeffers} {et~al.}(2018){Jeffers}, {Sch{\"o}fer}, {Lamert},
  {Reiners}, {Montes}, {Caballero}, {Cort{\'e}s-Contreras}, {Marvin},
  {Passegger}, {Zechmeister}, {Quirrenbach}, {Alonso-Floriano}, {Amado},
  {Bauer}, {Casal}, {Alonso}, {Herrero}, {Morales}, {Mundt}, {Ribas}, \&
  {Sarmiento}}]{Jeffers.etal:18}
{Jeffers}, S.~V., {Sch{\"o}fer}, P., {Lamert}, A., {et~al.} 2018, \aap, 614,
  A76

\bibitem[{{Jokipii}(1966)}]{Jokipii:66}
{Jokipii}, J.~R. 1966, \apj, 146, 480

\bibitem[{{Jokipii} \& {Coleman}(1968)}]{Jokipii.Coleman:68}
{Jokipii}, J.~R., \& {Coleman}, Jr., P.~J. 1968, \jgr, 73, 5495

\bibitem[{Kasting {et~al.}(1993)Kasting, Whitmire, \&
  Reynolds}]{Kasting.etal:93}
Kasting, J.~F., Whitmire, D.~P., \& Reynolds, R.~T. 1993, Icarus, 101, 108

\bibitem[{{Kay} {et~al.}(2016){Kay}, {Opher}, \& {Kornbleuth}}]{Kay.etal:16}
{Kay}, C., {Opher}, M., \& {Kornbleuth}, M. 2016, \apj, 826, 195

\bibitem[{{Kerwin} \& {Remmele}(2007)}]{Kerwin.Remmele:07}
{Kerwin}, B.~A., \& {Remmele}, R.~L. 2007, Journal of Pharmaceutical Sciences,
  96, 1468

\bibitem[{{Kopparapu} {et~al.}(2014){Kopparapu}, {Ramirez}, {SchottelKotte},
  {Kasting}, {Domagal-Goldman}, \& {Eymet}}]{Kopparapu.etal:14}
{Kopparapu}, R.~K., {Ramirez}, R.~M., {SchottelKotte}, J., {et~al.} 2014,
  \apjl, 787, L29

\bibitem[{{Laitinen} {et~al.}(2013){Laitinen}, {Dalla}, {Kelly}, \&
  {Marsh}}]{Laitinen.etal:13}
{Laitinen}, T., {Dalla}, S., {Kelly}, J., \& {Marsh}, M. 2013, \apj, 764, 168

\bibitem[{{Lingam} {et~al.}(2018){Lingam}, {Dong}, {Fang}, {Jakosky}, \&
  {Loeb}}]{Lingam.etal:18}
{Lingam}, M., {Dong}, C., {Fang}, X., {Jakosky}, B.~M., \& {Loeb}, A. 2018,
  \apj, 853, 10

\bibitem[{{Lingam} \& {Loeb}(2018)}]{Lingam.Loeb:18a}
{Lingam}, M., \& {Loeb}, A. 2018, ArXiv e-prints, arXiv:1810.02007

\bibitem[{{Loyd} {et~al.}(2018){Loyd}, {France}, {Youngblood}, {Schneider},
  {Brown}, {Hu}, {Segura}, {Linsky}, {Redfield}, {Tian}, {Rugheimer}, {Miguel},
  \& {Froning}}]{Loyd.etal:18}
{Loyd}, R.~O.~P., {France}, K., {Youngblood}, A., {et~al.} 2018, \apj, 867, 71

\bibitem[{{Luger} {et~al.}(2017){Luger}, {Sestovic}, {Kruse}, {Grimm},
  {Demory}, {Agol}, {Bolmont}, {Fabrycky}, {Fernandes}, {Van Grootel},
  {Burgasser}, {Gillon}, {Ingalls}, {Jehin}, {Raymond}, {Selsis}, {Triaud},
  {Barclay}, {Barentsen}, {Howell}, {Delrez}, {de Wit}, {Foreman-Mackey},
  {Holdsworth}, {Leconte}, {Lederer}, {Turbet}, {Almleaky}, {Benkhaldoun},
  {Magain}, {Morris}, {Heng}, \& {Queloz}}]{Luger.etal:17}
{Luger}, R., {Sestovic}, M., {Kruse}, E., {et~al.} 2017, Nature Astronomy, 1,
  0129

\bibitem[{{Mewaldt}(2006)}]{Mewaldt:06}
{Mewaldt}, R.~A. 2006, \ssr, 124, 303

\bibitem[{{Mewaldt} {et~al.}(2007){Mewaldt}, {Cohen}, {Haggerty}, {Mason},
  {Looper}, {von Rosenvinge}, \& {Wiedenbeck}}]{Mewaldt.etal:07}
{Mewaldt}, R.~A., {Cohen}, C.~M.~S., {Haggerty}, D.~K., {et~al.} 2007, in
  American Institute of Physics Conference Series, Vol. 932, Turbulence and
  Nonlinear Processes in Astrophysical Plasmas, ed. D.~{Shaikh} \& G.~P.
  {Zank}, 277--282

\bibitem[{{Mewaldt} {et~al.}(2008){Mewaldt}, {Cohen}, {Giacalone}, {Mason},
  {Chollet}, {Desai}, {Haggerty}, {Looper}, {Selesnick}, \&
  {Vourlidas}}]{Mewaldt.etal:08}
{Mewaldt}, R.~A., {Cohen}, C.~M.~S., {Giacalone}, J., {et~al.} 2008, in
  American Institute of Physics Conference Series, Vol. 1039, American
  Institute of Physics Conference Series, ed. G.~{Li}, Q.~{Hu},
  O.~{Verkhoglyadova}, G.~P. {Zank}, R.~P. {Lin}, \& J.~{Luhmann}, 111--117

\bibitem[{{Morin} {et~al.}(2010){Morin}, {Donati}, {Petit}, {Delfosse},
  {Forveille}, \& {Jardine}}]{Morin.etal:10}
{Morin}, J., {Donati}, J.-F., {Petit}, P., {et~al.} 2010, \mnras, 407, 2269

\bibitem[{{Osten} \& {Wolk}(2015)}]{Osten.Wolk:15}
{Osten}, R.~A., \& {Wolk}, S.~J. 2015, \apj, 809, 79

\bibitem[{{Powell} {et~al.}(1999){Powell}, {Roe}, {Linde}, {Gombosi}, \& {De
  Zeeuw}}]{Powell.etal:99}
{Powell}, K.~G., {Roe}, P.~L., {Linde}, T.~J., {Gombosi}, T.~I., \& {De Zeeuw},
  D.~L. 1999, Journal of Computational Physics, 154, 284

\bibitem[{{Reiners} \& {Basri}(2010)}]{Reiners.Basri:10}
{Reiners}, A., \& {Basri}, G. 2010, \apj, 710, 924

\bibitem[{{R{\'e}ville} {et~al.}(2015){R{\'e}ville}, {Brun}, {Matt},
  {Strugarek}, \& {Pinto}}]{Reville.etal:15}
{R{\'e}ville}, V., {Brun}, A.~S., {Matt}, S.~P., {Strugarek}, A., \& {Pinto},
  R.~F. 2015, \apj, 798, 116

\bibitem[{{Ribas} {et~al.}(2016){Ribas}, {Bolmont}, {Selsis}, {Reiners},
  {Leconte}, {Raymond}, {Engle}, {Guinan}, {Morin}, {Turbet}, {Forget}, \&
  {Anglada-Escud{\'e}}}]{Ribas.etal:16}
{Ribas}, I., {Bolmont}, E., {Selsis}, F., {et~al.} 2016, \aap, 596, A111

\bibitem[{{Segura} {et~al.}(2010){Segura}, {Walkowicz}, {Meadows}, {Kasting},
  \& {Hawley}}]{Segura.etal:10}
{Segura}, A., {Walkowicz}, L.~M., {Meadows}, V., {Kasting}, J., \& {Hawley}, S.
  2010, Astrobiology, 10, 751

\bibitem[{{Strauss} {et~al.}(2017){Strauss}, {Dresing}, \&
  {Engelbrecht}}]{Strauss.etal:17}
{Strauss}, R.~D.~T., {Dresing}, N., \& {Engelbrecht}, N.~E. 2017, \apj, 837, 43

\bibitem[{{Tilley} {et~al.}(2017){Tilley}, {Segura}, {Meadows}, {Hawley}, \&
  {Davenport}}]{Tilley.etal:17}
{Tilley}, M.~A., {Segura}, A., {Meadows}, V.~S., {Hawley}, S., \& {Davenport},
  J. 2017, arXiv e-prints, arXiv:1711.08484

\bibitem[{{T{\'o}th} {et~al.}(2012){T{\'o}th}, {van der Holst}, {Sokolov}, {De
  Zeeuw}, {Gombosi}, {Fang}, {Manchester}, {Meng}, {Najib}, {Powell}, {Stout},
  {Glocer}, {Ma}, \& {Opher}}]{Toth:12}
{T{\'o}th}, G., {van der Holst}, B., {Sokolov}, I.~V., {et~al.} 2012, Journal
  of Computational Physics, 231, 870

\bibitem[{{Turner} \& {Drake}(2009)}]{Turner.Drake:09}
{Turner}, N.~J., \& {Drake}, J.~F. 2009, \apj, 703, 2152

\bibitem[{{van der Holst} {et~al.}(2014){van der Holst}, {Sokolov}, {Meng},
  {Jin}, {Manchester}, {T{\'o}th}, \& {Gombosi}}]{vanderHolst:14}
{van der Holst}, B., {Sokolov}, I.~V., {Meng}, X., {et~al.} 2014, \apj, 782, 81

\bibitem[{{Veronig} {et~al.}(2002){Veronig}, {Temmer}, {Hanslmeier}, {Otruba},
  \& {Messerotti}}]{Veronig.etal:02}
{Veronig}, A., {Temmer}, M., {Hanslmeier}, A., {Otruba}, W., \& {Messerotti},
  M. 2002, \aap, 382, 1070

\bibitem[{{Vida} {et~al.}(2017){Vida}, {K{\H o}v{\'a}ri}, {P{\'a}l},
  {Ol{\'a}h}, \& {Kriskovics}}]{Vida.etal:17}
{Vida}, K., {K{\H o}v{\'a}ri}, Z., {P{\'a}l}, A., {Ol{\'a}h}, K., \&
  {Kriskovics}, L. 2017, \apj, 841, 124

\bibitem[{{Vidotto} {et~al.}(2014){Vidotto}, {Jardine}, {Morin}, {Donati},
  {Opher}, \& {Gombosi}}]{Vidotto.etal:14}
{Vidotto}, A.~A., {Jardine}, M., {Morin}, J., {et~al.} 2014, \mnras, 438, 1162

\bibitem[{{Youngblood} {et~al.}(2017){Youngblood}, {France}, {Loyd}, {Brown},
  {Mason}, {Schneider}, {Tilley}, {Berta-Thompson}, {Buccino}, {Froning},
  {Hawley}, {Linsky}, {Mauas}, {Redfield}, {Kowalski}, {Miguel}, {Newton},
  {Rugheimer}, {Segura}, {Roberge}, \& {Vieytes}}]{Youngblood.etal:17}
{Youngblood}, A., {France}, K., {Loyd}, R.~O.~P., {et~al.} 2017, \apj, 843, 31

\bibitem[{{Zhang} {et~al.}(2003){Zhang}, {McKibben}, {Lopate}, {Jokipii},
  {Giacalone}, {Kallenrode}, \& {Rassoul}}]{Zhang.etal:03}
{Zhang}, M., {McKibben}, R.~B., {Lopate}, C., {et~al.} 2003, Journal of
  Geophysical Research (Space Physics), 108, 1154

\end{thebibliography}



\end{document}